\documentclass[fleqn,usenatbib]{mnras}

\usepackage[T1]{fontenc}

\DeclareRobustCommand{\VAN}[3]{#2}
\let\VANthebibliography\thebibliography
\def\thebibliography{\DeclareRobustCommand{\VAN}[3]{##3}\VANthebibliography}

\usepackage{graphicx}	
\usepackage{amsmath}	
\usepackage{amssymb}	

\usepackage{newtxtext,newtxmath}  

\usepackage{color}

\title[Rotation rates in eccentric binaries]{Stellar rotation rates in {\it Kepler} eccentric (heartbeat) binaries obtained from r-mode signatures}

\author[H. Saio and D. W. Kurtz]{
Hideyuki Saio$^1$\thanks{E-mail: saio@astr.tohoku.ac.jp}
and Donald W. Kurtz$^{2,3}$
\\
$^{1}$ Astronomical Institute, Graduate School of Science, Tohoku University, Sendai, Miyagi 980-8578, Japan\\
$^{2}$Centre for Space Research, Physics Department, North-West University, Mahikeng 2745, South Africa\\
$^{3}$Jeremiah Horrocks Institute, University of Central Lancashire, Preston PR1 2HE, UK\\
}

\date{Accepted XXX. Received YYY; in original form ZZZ}

\pubyear{2021}

\begin{document}
\label{firstpage}
\pagerange{\pageref{firstpage}--\pageref{lastpage}}
\maketitle

\begin{abstract}
R-mode oscillations in a rotating star produce characteristic signatures in  a Fourier amplitude spectrum at frequencies related with the rotation frequency, which can be, in turn, used to obtain the surface rotation rate of the star. Some binary stars observed by {\it Kepler} indicate the presence of r~modes that are probably excited by the tidal effect. In this paper, we have obtained stellar rotation periods in 20 eccentric (heartbeat) binaries with r-mode signatures. The majority of the rotation periods are found to be comparable to pseudo-synchronous periods, in which the angular velocity of rotation is similar to the angular orbital motion of the companion at periastron. In particular, for the heartbeat stars with orbital periods longer than about 8\,d, all but one agree with pseudo-synchronous rotation. In contrast to a previous investigation by Zimmerman et al., our result supports the pseudo-synchronisation theory developed by Hut.
\end{abstract}




\section{Introduction}

Among close binaries there is a subclass having large eccentricities. The light curves of such highly eccentric binary systems exhibit rapid variations caused by strong tidal effects during periastron passage. These systems are called `heartbeat' stars \citep{Thompson2012} because of the resemblance between the light curve and an electrocardiograph. Thanks to the precise and nearly uninterrupted observations from the {\it Kepler} satellite, the number of the identified heartbeat stars has increased greatly as 172 systems are now listed in the {\it Kepler} Eclipsing Binary Catalogue  \citep[KEBC,][]{KEBC2016}. Heartbeat stars are also found in TESS data  \citep{Kolaczek2021,IJspeert2021}, and most recently more than 200 have been found (Hambleton, private communication), and 991 in OGLE data \citep{Wrona2021}.

The presence of many eccentric binaries was predicted by \citet{Hut1981}, who found, from a theoretical analysis, that if the ratio of the orbital to rotational angular momentum is sufficiently high ($\ga 7$), the timescale of decreasing the orbital eccentricity by tidal dissipation is much longer than the changes of other binary parameters. \citet{Hut1981} predicted the rotation rate in such a binary system to settle into pseudo-synchronisation with an eccentric orbit, in which the angular rotation velocity of the primary star should be close to (but slightly slower than) the orbital angular velocity of the secondary at the periastron (much faster than the mean motion).  Such eccentric binaries should correspond to heartbeat stars.    

An Fourier transform amplitude spectrum (FT hereafter) of the light curve of a heartbeat star reveals a long series of equally spaced orbital frequency harmonics caused by the strongly non-sinusoidal light curve. In addition, it often shows amplitudes at frequencies associated with stellar spots and pulsations (oscillations).

Assuming that a peak (or a tight group of peaks ) and harmonics separated from the orbital peaks in an FT are caused by a spot, or spots, on the rotating stellar surface, \citet{Zimmerman2017} obtained the rotation periods for stars in 24 heartbeat systems. For 18 of them with eccentricities evaluated in the literature, they compared the rotation periods obtained with pseudo-synchronous rotation periods that are determined from the orbital period and eccentricity \citep{Hut1981}. \citeauthor{Zimmerman2017} found that rotation periods of many heartbeat binaries are about $\frac{3}{2}$ times the pseudo-synchronous rotation periods, contrary to the prediction of \citet{Hut1981}.

Among the peaks that \citet{Zimmerman2017} adopted as rotational modulations  due to surface spots, there are cases in which the signature is not a single peak but a group of peaks spread over a certain range of frequencies. \citeauthor{Zimmerman2017} adopted the central frequency of the peak distribution as the rotation frequency,  assuming the group to be caused by spots on different latitudes in the presence of latitudinal differential rotation.

Instead, we assume in the present paper such a group of frequencies to be caused by r~modes (global Rossby waves influenced by buoyancy).  The strong tidal forces would disturb the rotational flow in the stellar envelope, which would, in turn, generate Rossby waves and hence their global modes, i.e., r~modes, would be formed as discussed in \citet{Saio+2018} and \citet{Saio2018}. 
The presence of r modes in binary stars is not limited to highly eccentric binaries like heartbeat stars, but is common in close binaries associated with strong tidal forces. For example, \cite{Saio2020} found r-mode signatures in $\sim$\!700 stars among $\sim$\!800 eclipsing binary samples of the KEBC with orbital periods between 0.4 and 5 d and having {\it Kepler} light curves longer than 3 quarters. 

The r-mode features in an FT of the light curve of a star are useful to estimate the rotation rate as discussed in \citet{Saio2019,Saio2020}. Thus, we obtain rotation periods by fitting r~modes with expected visibility distributions for twenty heartbeat stars having r-mode signatures selected from those analysed by \citet{Zimmerman2017} and additional heartbeat stars whose orbital eccentricities are obtained in the literature. 

\section{Method of analysis}

\subsection{{\it Kepler} long-cadence data to Fourier amplitude spectra}

{\it Kepler} long-cadence photometric data of heartbeat stars were downloaded from the KASOC (Kepler Asteroseismic Science Operations Center) website\footnote{https://kasoc.phys.au.dk/}. Flux data (corrected by KASOC) for each {\it Kepler} Quarter were converted to magnitude, and  the mean of the data for each Quarter was subtracted from the original data. Then, using the software Period04 \citep{Period04} Fourier analyses were performed for the combined data for all the available {\it Kepler} Quarters for each binary system.\footnote{For KIC\,3230227 and KIC\,11403032, the de-trended long-cadence data were downloaded from KEBC, because lower noise levels in the FTs were obtained. } We use amplitude spectra, which are referred to hereafter as FTs (for Fourier transforms). All pre-whitening to produce FTs of the residuals after removing orbital harmonic series was done using Period04. All of our FTs are in the low-frequency range where r~modes are visible. The long-cadence {\it Kepler} data has a Nyquist frequency a bit greater than 24\,d$^{-1}$, but we do not show the higher frequency ranges as they are not relevant to the discussion in this paper. 

\begin{figure}
\includegraphics[width=0.49\textwidth]{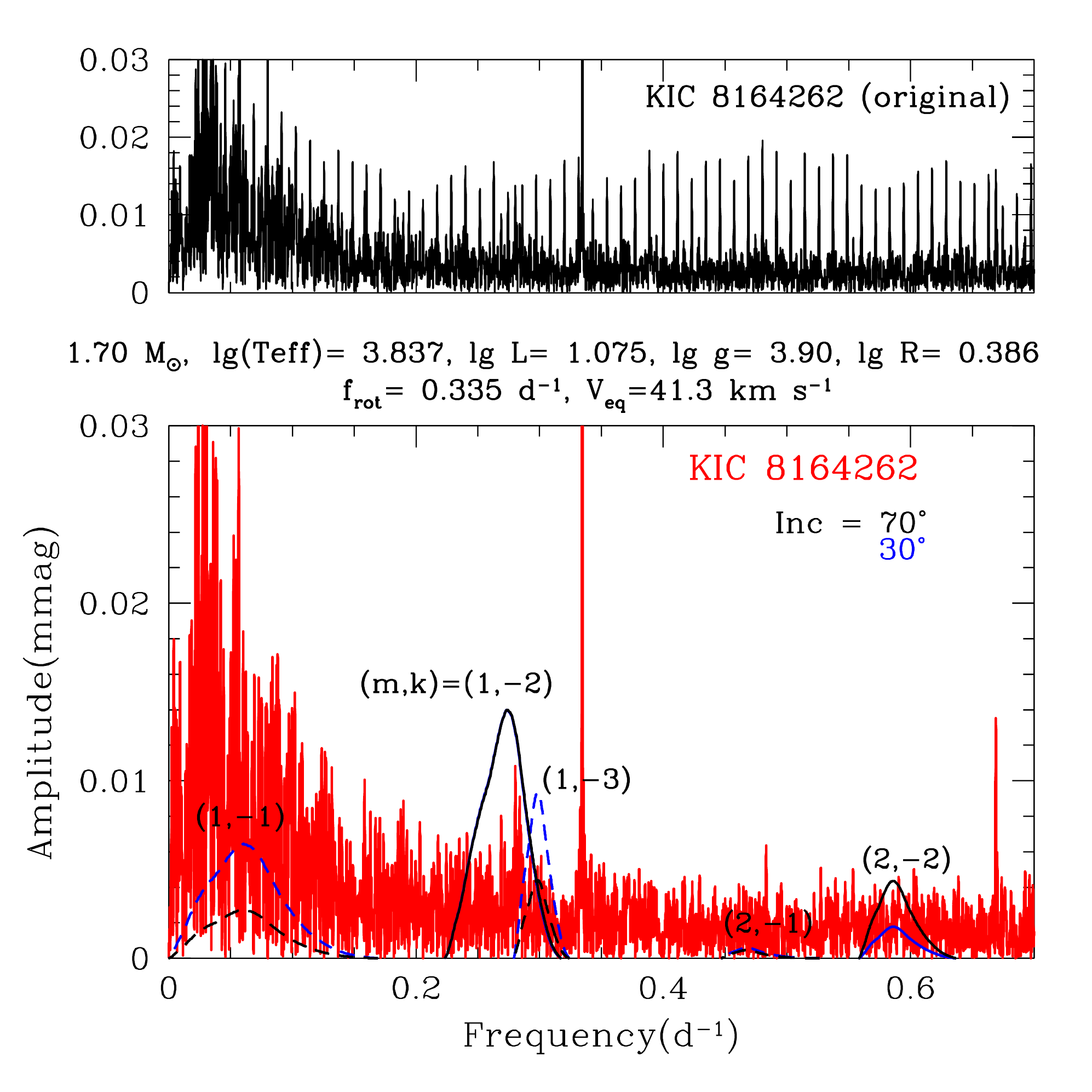}
\caption{FTs for KIC\,8164262 and r-mode visibilities at a rotation frequency of $0.335$\,d$^{-1}$ for a 1.70-M$_\odot$ model. 
The upper panel shows the FT from the original light curve, while red line in the lower panel shows the FT after the orbital harmonics have been removed. The predicted visibility distributions are shown by solid and dashed lines for even  $(k=-2)$ and odd $(k=-1,-3)$ modes, respectively. Black and blue lines are for inclinations of $70^\circ$ and $30^\circ$, respectively. They are arbitrarily normalised to 0.014\,mmag at the maximum visibility for $(m,k)=(1,-2)$, for which the blue and black lines are identical. The peaks at 0.335\,d$^{-1}$ and 0.670\,d$^{-1}$ are the rotation frequency and its harmonic.
}
\label{fig:k816}
\end{figure}

We explain how we proceeded with our analysis with the example of the heartbeat star KIC\,8164262, on which \citet{Hambleton2018} performed detailed analysis. KIC\,8164262 has a relatively long orbital period of 87.45717\,d (from the KEBC) and a large eccentricity of $0.886$ \citep{Hambleton2018}. Fig.\,\ref{fig:k816} shows FTs for KIC\,8164262. The upper panel shows the FT of the original light curve. The equally spaced (by $f_{\rm orb}=0.01434$\,d$^{-1}$) lines at $f_{\rm orb}$ and many harmonics are caused by the rapid (strongly non-sinusoidal) light variation during periastron.

The orbital effects were removed by using Period04~\citep{Period04},  in which the orbital effects were expressed as 
 a sum of $jmax$ sinusoidal variations  
\begin{equation}
\sum_{j=1}^{jmax} A_j\sin[2\uppi(jf_{\rm orb}(t-t_0)+\phi_j)],
\label{eq:jmax}
\end{equation}
with amplitude $A_j$ and phase $\phi_j$, where $t$ and $t_0$ are time in days and a fiducial one, respectively.  Period04 determines iteratively the best set of $(A_j,\phi_j)$ for $1 \le j \le jmax$ to minimise the residuals. For KIC\,8164262 we have included 100 harmonics ($jmax=100$)\footnote{
For each case we adopted $jmax$ large enough that no harmonic of significant amplitude was left in the FT, or that any such harmonic was at far higher frequency than the r modes so did not affect the analysis of those.}. 
The sum of the sinusoidal series was subtracted from the original data. The FT of the residual data is shown by the red line in Fig.\,\ref{fig:k816}. In accordance with the results of \citet{Hambleton2018}, our FT shows small peaks around $0.28$\,d$^{-1}$ as well as peaks at the rotational frequency $0.335$\,d$^{-1}$ and its harmonic. 

\subsection{R-mode frequency ranges}\label{sec:freqrange}

Although \citet{Hambleton2018} identified the frequencies of KIC\,8164262 around $0.28$\,d$^{-1}$ as g~modes (retrograde to the rotation), we regard these frequencies, which are lower than the rotation frequency, as r~modes. Those oscillations are global modes of Rossby waves influenced by buoyancy, whose frequency range expected in the observer's frame is given as
\begin{equation}
mf_{\rm rot}\left[1-{2\over(m+|k|-1)(m+|k|)}\right] < \nu(m,k,n) < mf_{\rm rot} 
\label{eq:freqrange}
\end{equation}
with azimuthal order $m$ ($>0$ for r~modes; in this paper we adopt the convention that $m>0$ corresponds to retrograde modes in the co-rotating frame), latitudinal order $k$ ($\le -1$ for r~modes), and radial order $n$ ($>0$)  \citep[see e.g.][for details]{Saio+2018}. Even $|k|$ corresponds to a pulsation mode whose light variation on the surface is symmetric with respect to the equator, while odd $|k|$ corresponds to an  anti-symmetric one. The number of latitudinal nodal lines are $|k+2|$ for even r~modes and $|k|$ for odd r~modes.

In the co-rotating frame, the r-mode frequency, $\nu^{\rm co}(m,k,n)$, decreases with increasing radial order $n$, as for g~modes. 
However, since the frequency in the observer's frame is given as $\nu(m,k,n) = |\nu^{\rm co}(m,k,n) - mf_{\rm rot}|$ and $\nu^{\rm co}(m,k,n) < mf_{\rm rot}$, $\nu(m,k,n)$ increases (getting closer to $mf_{\rm rot}$) with increasing $n$. For this reason, the period spacing of r~modes increases with increasing period, which is opposite to the period spacings of prograde g~modes, as is well known in $\gamma$\,Dor stars \citep[e.g.][]{VanReeth2016,LiG2019}. 

If we assume that all of the r~modes have equal kinetic energy, we can calculate the visibility distribution as a function of pulsation frequency, by integrating the pulsation amplitude across the stellar surface \citep[see][for details]{Saio+2018}. Examples of the visibility distribution for a model of  KIC\,8164262 with $f_{\rm rot}= 0.335$\,d$^{-1}$ are shown in Fig.\,\ref{fig:k816} (lower panel), in which the solid and dashed curves are for even ($k=-2$) and odd ($k=-1, -3$) modes, respectively. Unsurprisingly, odd mode visibility is smaller for a larger inclination angle. In this figure, r-mode frequency ranges have values less than $f_{\rm rot}$ for $m=1$ r~modes, and are between $f_{\rm rot}$ and $2f_{\rm rot}$ for $m=2$ r~modes. 

The most visible feature of r~modes in an FT is a frequency group observed somewhere between $\frac{2}{3}f_{\rm rot}$ and $f_{\rm rot}$; i.e. even r~modes of $(m,k)=(1,-2)$.  This is a general property of r modes as discussed in \S3 of \citet{Saio+2018}.  The expected visibility distribution is normalised arbitrarily at the maximum of the $(m,k)=(1,-2)$ mode visibility for all the cases in this paper. 
The frequency at the maximum visibility within the frequency range depends on the rotation frequency; it is close to $\frac{2}{3}f_{\rm rot}$ for slow rotations (say $f_{\rm rot}< 0.2$~d$^{-1}$), while it  increases to $\sim$$0.9f_{\rm rot}$ for $f_{\rm rot}= 0.6$~d$^{-1}$ in our sample. We try to fit both maximum visibility and frequency range. If it is not possible, we give priority to fit the visibility range assuming that the frequency at maximum visibility may be affected by the tidal effect which is not included. 

Fig.\,\ref{fig:k816} shows that the $\sim$$0.28$\,d$^{-1}$ frequencies are consistent with the visibility distribution for r~modes of $(m,k)=(1,-2)$ for the rotation frequency $f_{\rm rot}= 0.335$\,d$^{-1}$, which itself agrees with the frequency at one of the peaks generated by a stellar spot. In other words, by fitting a frequency group with the visibility distribution of r~modes, we can obtain the rotation rate even if in the FT no peak caused by spots is present.  More discussion on KIC\,8164246 will be given in \S\ref{sec:k816} below.

\subsection{Pseudo-synchronous rotation periods}

\cite{Hut1981} found that in an eccentric binary with orbital angular momentum that is much larger than the rotational angular momentum, the orbital motion and stellar rotation attain pseudo-synchronisation much faster than attaining synchronous circularisation.  \citet{Hut1981} derived an analytic relation among the  pseudo-synchronous rotation period $P_{\rm ps-rot}$, the orbital eccentricity $e$, and the orbital period $P_{\rm orb}$ as 
\begin{equation}
    P_{\rm ps-rot}=\frac{(1+3e^2+{3\over8}e^4)(1-e^2)^{3/2}}{1+{15\over2}e^2+{45\over8}e^4+{5\over16}e^6}P_{\rm orb}.
	\label{eq:ps}
\end{equation}
Using parameters of KIC\,8164262, $e = 0.886$ \citep{Hambleton2018} and $P_{\rm orb}=87.457$\,d in the above equation, we obtain $P_{\rm ps-rot} = 2.976$\,d, while $f_{\rm rot}= 0.335$\,d$^{-1}$ corresponds to a rotation period of $2.985$\,d which is very close to $P_{\rm ps-rot}$, indicating the system is pseudo-synchronous. We note, however, that $P_{\rm ps-rot}$ is sometimes very sensitive to the eccentricity, $e$, because equation (\ref{eq:ps}) involves high order terms of $e$; for example, \citet{Zimmerman2017} obtained $P_{\rm ps-rot}=4.22$\,d for KIC\,8164262, using $e=0.857$ \citep{Shporer2016}.
\footnote{
In this paper, we adopt $e = 0.886$ obtained by \citet{Hambleton2018} for KIC\,8164262 rather than $e=0.857$ obtained by \citet{Shporer2016}, because more radial-velocity measurements were incorporated in the former analysis.}

In this paper, we select twenty heartbeat binaries which show r-mode frequency groups in the FT,  and whose orbital eccentricities  are obtained in the literature. Comparing these r-mode features with expected r-mode frequency ranges for assumed rotation rates we determine the rotation frequency of the primary star and compare it with the corresponding $P_{\rm ps-rot}$. Model parameters for each case are selected mainly referring to the parameters given in \citet{Berger2020}; we have chosen a mass taking into account the luminosity, then chosen an evolutionary stage which is consistent with the effective temperature.  Model parameters need not be very accurate because the predicted visibility distribution of r~modes for a given rotation frequency is only weakly dependent on the stellar parameters, as is the rotation rate derived by fitting. Stellar~models were obtained, as in our previous models for $\gamma$\,Dor stars \citep{Saio2021}, using the MESA code \citep[ver.7184][]{pax11,pax13,pax15} adopting the same settings. In our present paper we adopt models with diffusive overshooting of $h_{\rm os}=0.01$, in which mixing at distance $z$ from the convective-core boundary is  proportional to $\exp[-2z/(h_{\rm os}H_p)]$ with pressure scale height $H_p$.  

\section{R-mode fits for some heartbeat stars}

In this section, we show the fitting of frequency groups with r~mode visibility for four selected cases, while the rest are shown in the Appendix.
Fitting results and model parameters are summarised in Table~\ref{tab:sum}. 

\subsection{KIC\,8164262}\label{sec:k816}

\citet{Hambleton2018} carried out a detailed analysis of the heartbeat star KIC\,8164262 to obtain the binary orbital parameters as well as stellar parameters of the primary and secondary components. They found many tidally excited oscillations whose frequencies are multiples of the orbital frequency. They also found a low frequency, $0.335$\,d$^{-1}$, with its harmonic, which they identified as the rotation frequency, and three low-amplitude frequencies around $0.28$\,d$^{-1}$, which they regarded as retrograde g~modes. 

Adopting $0.335$\,d$^{-1}$ as the rotation frequency for a 1.70-${\rm M}_\odot$ model with $(\log T_{\rm eff}, \log R)=(3.837,0.386)$, which are consistent with the parameters obtained by \citet{Hambleton2018}, we obtained r-mode visibility for $m=1$ and 2 (Fig.\,\ref{fig:k816}).  The three frequencies found by \citet{Hambleton2018} around $0.28$\,d$^{-1}$  are located within the frequency range of r~modes with $(m,k)=(1,-2)$. 

Although not obvious in this figure, a small hump caused by r~modes with $(m,k)=(2,-2)$ is visible around $0.6$\,d$^{-1}$ in Figure\,7 of \citet{Hambleton2018}, which confirms the r-mode interpretation for the humps in KIC\,8164262. As mentioned above, the eccentricity $0.866\pm 0.003$ obtained by \citet{Hambleton2018} yields a pseudo-synchronous rotation period of  $2.98\pm0.12$\,d, which is consistent with the rotation period $2.985$\,d.  The rotation period with the radius of the model corresponds to an equatorial rotation velocity of $41$~km~s$^{-1}$. Combining this velocity with $v\sin i = 23\pm1$~km~s$^{-1}$ obtained by \citet{Hambleton2018} yields an inclination angle of $34^\circ$ between the rotation axis and the line-of-sight, which is, in turn, inclined to the orbital axis by about $30^\circ$ \citep{Hambleton2018}. The misalignment between the rotation and the orbital axes is possible in the binary evolution theory of \citet{Hut1981}, in which the time scale of the alignment   is comparable to or slightly longer than the time scale of the pseudo-synchronization \citep[see Fig. 4 of][]{Hut1981}.   

\subsection{KIC\,9899216}\label{sec:k989}

\begin{figure}
\includegraphics[width=0.49\textwidth]{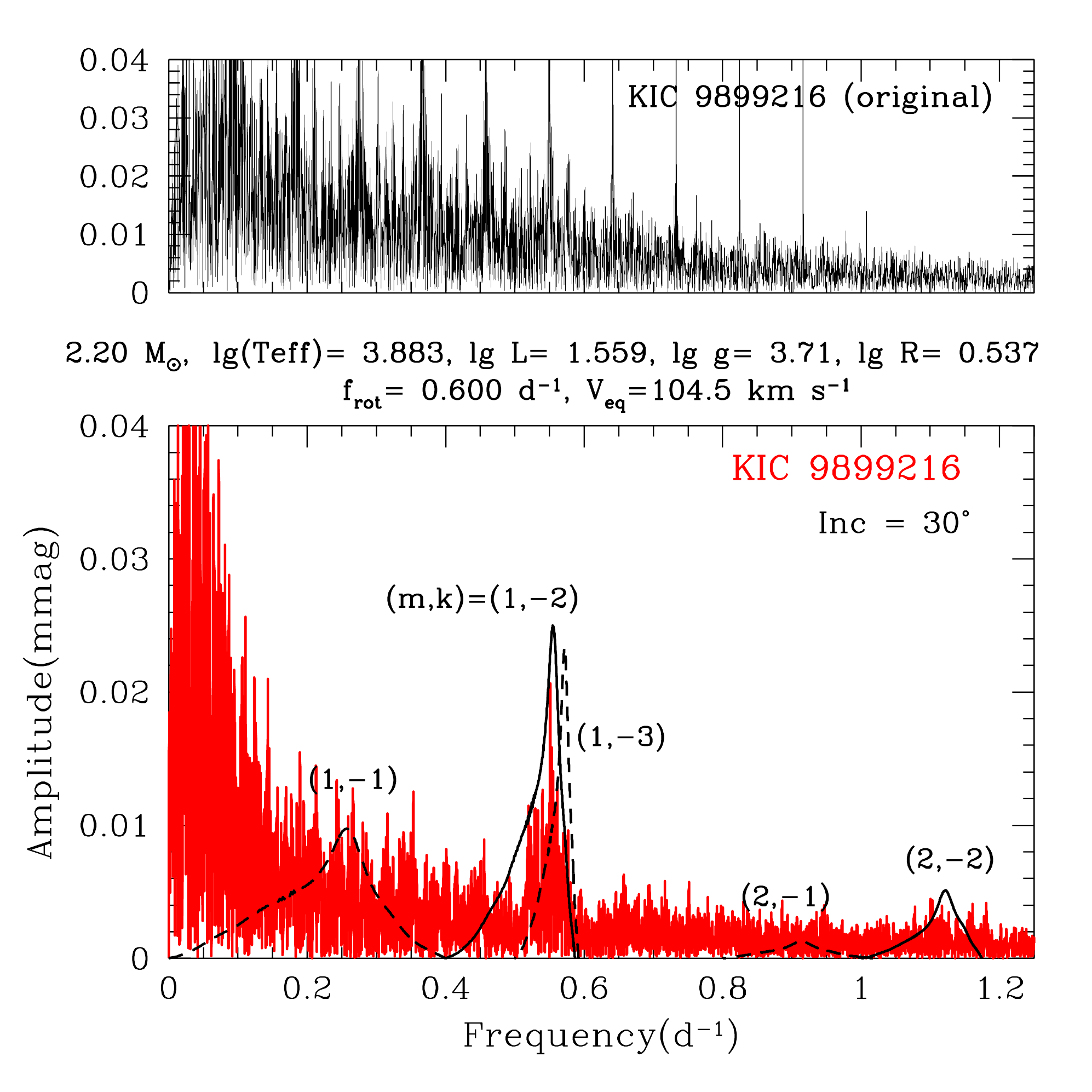}
\caption{FTs for KIC\,9899216 shown in the same format as in Fig.\,\ref{fig:k816}. R-mode visibilities for a low-inclination case are shown because \citet{Thompson2012} obtained an orbital inclination of $20.7\pm 0.3~^{\circ}$, while \citet{Smullen2015} estimated $6-15~^\circ$.}
\label{fig:k989}
\end{figure}

The orbital period of the heartbeat binary KIC\,9899216 is $10.916$\,d (KEBC). Fig.\,\ref{fig:k989} shows FTs for KIC\,9899216. The FT derived from the original data (upper panel) is dominated by features caused by light variations associated with the orbital motion.  After pre-whitening 40 orbital harmonics from the original data, we obtain the FT  of the residual data that is shown by the red line in the lower panel of Fig.\,\ref{fig:k989}; a clear frequency group has emerged around $0.55$\,d$^{-1}$ ($\approx 6/P_{\rm orb}$). Its frequency range is fitted with r~modes of $(m,k)=(1,-2)$ at a rotation frequency of $0.60$\,d$^{-1}$ ($P_{\rm rot}= 1.67$\,d).  This is the fastest rotation among our sample of heartbeat stars.

The orbital eccentricity of KIC\,9899216 was determined from the {\it Kepler} light curve data by \citet{Thompson2012}, who obtained $e=0.647\pm0.004$, while \citet{Smullen2015} obtained $e=0.66\pm0.24$ analysing their spectroscopic radial-velocity measurements.  In Table~\ref{tab:sum}, we have adopted $e=0.65\pm0.01$. Using the eccentricity and the orbital period $10.916$\,d in equation (\ref{eq:ps}), we obtain $P_{\rm ps-rot}= 2.15\pm0.10$\,d, which is slightly larger than the rotation period obtained above.

\subsection{KIC\,3230227 (HD 181850)}\label{sec:k323}

KIC\,3230227 is a heartbeat binary with an orbital period of $7.0471$\,d. A very narrow eclipse is seen in the folded {\it Kepler} light curve \citep[e.g., KEBC,][]{KEBC2016}. Recent orbital analyses by three groups for KIC\,3230227 using the {\it Kepler} light curve and spectroscopic velocity observations have yielded similar results; \citet{Dimitrov2017} obtained an orbital eccentricity of $e=0.603\pm0.001$ and an orbital inclination $i= 72.8 \pm 0.2^{\circ}$,  \citet{Guo2017} obtained $e= 0.600 \pm 0.005$, $i=73.4 \pm 0.3^{\circ}$, and \citet{Smullen2015} obtained $e= 0.60\pm 0.04$, $i=66-71^{\circ}$ but with a slightly larger orbital period $7.051$\,d in the last case. Adopting $e=0.602\pm 0.003$ and $P_{\rm orb}=7.0471$\,d in equation (\ref{eq:ps}) yields  $P_{\rm ps-rot}=1.71\pm0.02$\,d as listed in Table~\ref{tab:sum}.  

\begin{figure}
\includegraphics[width=0.49\textwidth]{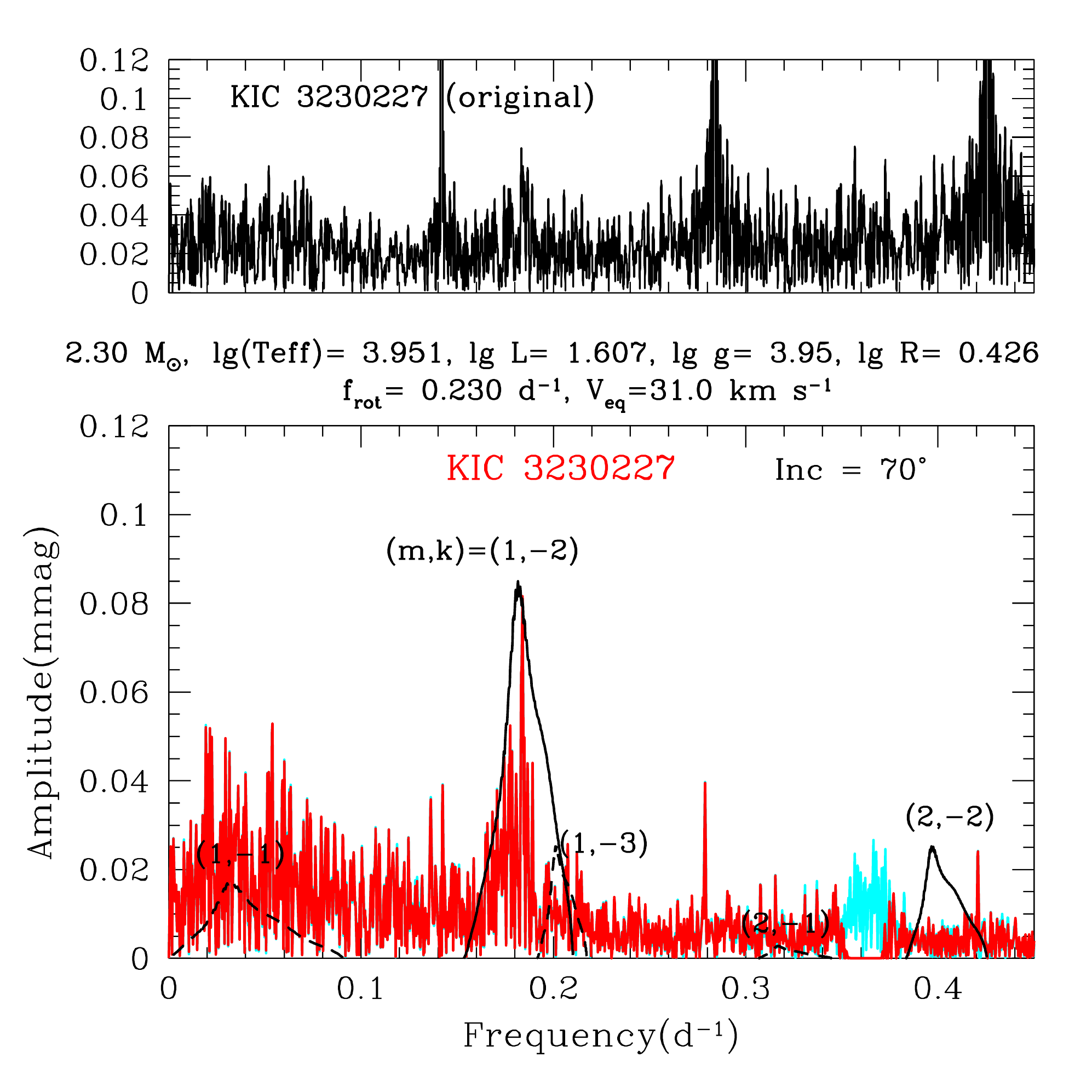}
\caption{FTs for KIC\,3230227 and theoretical visibility distributions of a model at a rotation frequency of $0.23$\,d$^{-1}$. The upper panel shows the FT for the original light curve including orbital effects. Removing 50 orbital harmonics, we obtain the red and cyan lines in the lower panel. The cyan line indicates frequencies corresponding to harmonics and combination frequencies of peaks around $0.18$\,d$^{-1}$ ($>0.04$~mmag). This indicates the secondary 'hump' at $\sim\!0.36\,$d$^{-1}$ need not to be considered as $m=2$ r modes. 
}
\label{fig:k323}
\end{figure}

Fig.\,\ref{fig:k323} shows FTs for KIC\,3230227; one for the original data (upper panel) and one for the residuals after removing the orbital harmonics (lower panel). An r-mode visibility distribution for a 2.30-M$_\odot$ model at a rotation frequency of $0.23$\,d$^{-1}$ is shown in a similar way to Fig.\,\ref{fig:k816}. This star has the main group of peaks at $\sim$$0.18$\,d$^{-1}$ and, as for many other cases, a secondary group of amplitudes that is formed by harmonics and combination frequencies among the main group of frequencies, which are shown by cyan lines in Fig.\,\ref{fig:k323}. Ignoring the secondary group, we consider only the main group. As shown in Fig.\,\ref{fig:k323}, the main frequency group around $0.18$\,d$^{-1}$ is fitted with $(m,k)=(1,-2)$ r~modes at a rotation frequency of $0.23$\,d$^{-1}$. The rotation frequency corresponds to  $P_{\rm rot}=4.35$\,d, which is larger than the $P_{\rm ps-rot}$ obtained above for KIC\,3230227. This deviation from the pseudo-synchronous rotation tends to occur in systems with orbital periods shorter than 8\,d (see Table\,\ref{tab:sum} and Fig.\,\ref{fig:rot_Porb} below).

The model shown in Fig.\,\ref{fig:k323} has an equatorial rotational velocity of $31$~km~s$^{-1}$, which agrees with  $v\sin i \approx30$~km~s$^{-1}$ obtained for the primary star of KIC\,3230227 by \citet{Smullen2015} ($\approx75$~km~s$^{-1}$ for the secondary).

\subsection{KIC\,5017127}\label{sec:k501}

\begin{figure}
\includegraphics[width=0.49\textwidth]{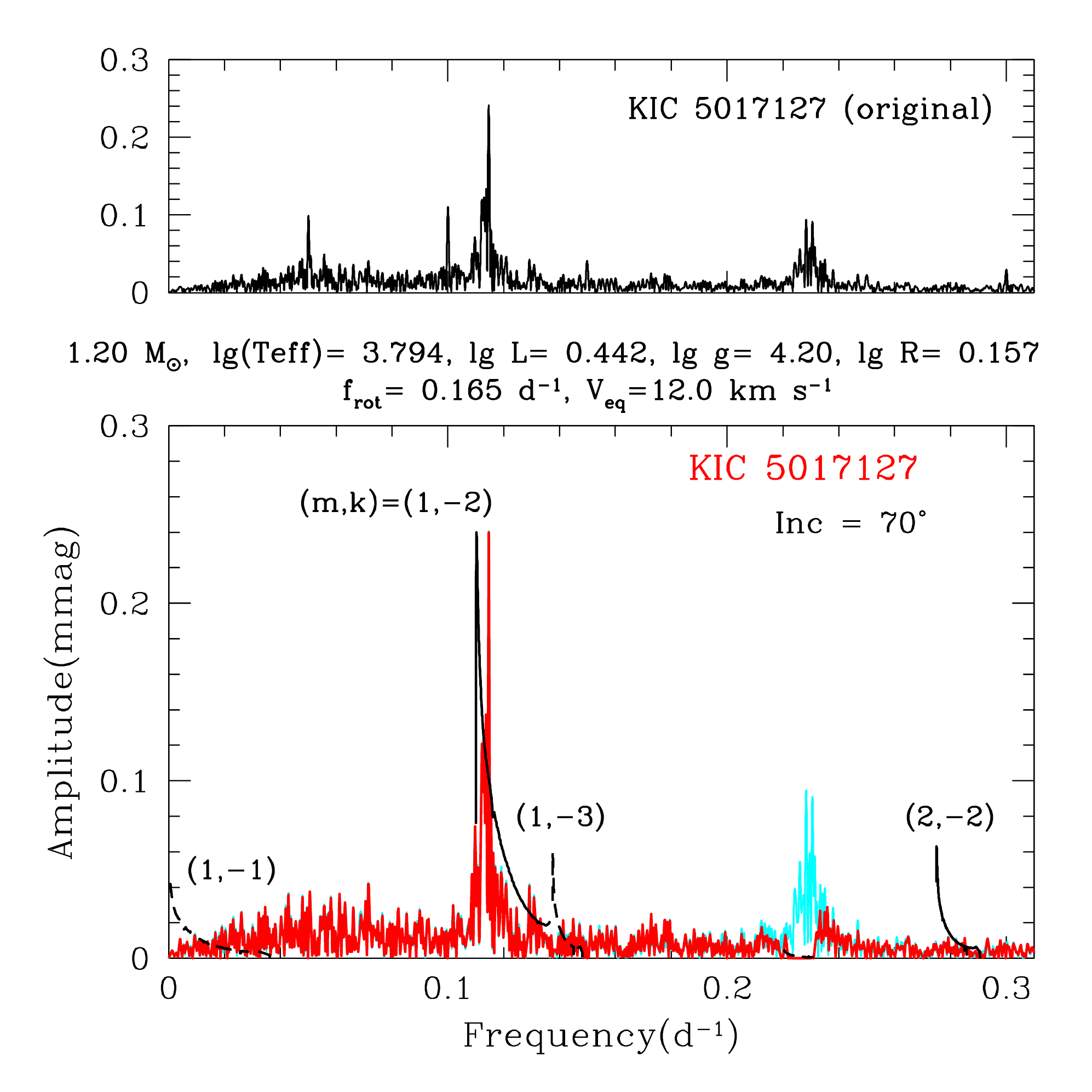}
\caption{FTs for KIC\,5017127. The upper panel shows an FT of the original light curve including the orbital harmonics from the non-sinusoidal light curve. The lower panel shows the FT after pre-whitening the orbital harmonics. The cyan line denotes harmonics and combination frequencies of the six highest amplitude ($>0.07$~mmag) peaks in the group at $\sim$$0.11$\,d$^{-1}$. The main group is fitted with the visibility distribution of $(m,k)=(1,-2)$ r~modes in a 1.2-M$_\odot$ model with a rotation frequency of $0.165$\,d$^{-1}$. The predicted frequency range for the low mass and slow rotation model is very narrow, which is consistent with the observed amplitude group.}
\label{fig:k501}
\end{figure}

The orbital period of KIC\,5017127 is $20.006$\,d (KEBC) with the eccentricity  $0.550\pm0.005$ \citep{Shporer2016}, from which we obtain $P_{\rm ps-rot}=5.97\pm0.05$\,d (eq.\,\ref{eq:ps}). An FT of KIC\,5017127 (Fig.\,\ref{fig:k501}) shows a narrow group of peaks at a frequency of $0.11$\,d$^{-1}$, which \citet{Zimmerman2017} identified as the rotation frequency of the primary and obtained $P_{\rm rot}=9.34\pm0.11$\,d. Instead, we fit the frequency group  with $(m,k)=(1,-2)$ r~modes at a rotation frequency of $0.165$\,d$^{-1}$ ($P_{\rm rot}= 6.06$\,d) (Fig.\,\ref{fig:k501}). The rotation period obtained is similar to the pseudo-synchronous rotation period, while it is smaller than the result of \citet{Zimmerman2017} by a factor of $\sim$$1.5$. 

\section{Summary of the results}

\begin{table*}
	\centering
	\caption{Observational and model parameters, and obtained rotation periods. The orbital period $P_{\rm orb}$ of each binary is taken from the KEBC \citep[Kepler Eclipsing Binary Catalog;][http://keplerEBs.villanova.edu]{KEBC2016}. The values in parentheses denote the uncertainty of the last digits.}
	\label{tab:sum}
	\begin{tabular}{rrrrrrrrrrrrc} %
		\hline
  & \multicolumn{3}{c}{\citet{Berger2020}} &  & & &
  \multicolumn{3}{c}{Model parameters} \\
   \cline{2-4}  \cline{8-10}\\
  \multicolumn{1}{c}{KIC} & \multicolumn{1}{c}{$T_{\rm eff}$} & \multicolumn{1}{c}{$\lg g$} & \multicolumn{1}{c}{$\lg L$}  & \multicolumn{1}{c}{$P_{\rm orb}$} & \multicolumn{1}{c}{eccentricity} & \multicolumn{1}{c}{$P_{\rm ps-rot}$} & \multicolumn{1}{c}{$M$} & \multicolumn{1}{c}{$\lg T_{\rm eff}$} & \multicolumn{1}{c}{$\lg L$} & \multicolumn{1}{c}{$P_{\rm rot}$} & \multicolumn{1}{c}{$V_{\rm eq}$} & \multicolumn{1}{c}{ref~$^{\rm a}$} \\
 & \multicolumn{1}{c}{($10^3$K)} & \multicolumn{1}{c}{(cgs)} & \multicolumn{1}{c}{(L$_\odot$)} &  \multicolumn{1}{c}{(d)} & & \multicolumn{1}{c}{(d)} & \multicolumn{1}{c}{(${\rm M}_\odot$)} & \multicolumn{1}{c}{(K)} & (L$_\odot$)  & \multicolumn{1}{c}{(d)} & \multicolumn{1}{c}{(km~s$^{-1}$)} \\
		\hline
 3230227 & 9.11(30) & 3.94(4) & 1.64(69) & 7.0471 & 0.602(3) & 1.71(2) & 2.30 & 3.951 & 1.607  & 4.35(10) & 31 & d17,g17 \\ 
 4372379 & 6.73(14) & 4.23(3) & 0.61(54) & 4.5352 & 0.342(92) & 2.61(64) & 1.40 & 3.828 & 0.678 & 4.76(24) & 17 & t12 \\
 5017127 & 6.29(14) & 4.30(11) & 0.36(14) & 20.006 & 0.550(5) & 5.97(11) & 1.20 & 3.794 & 0.442 & 6.06(19) & 12 & s16 \\
 5034333 & 9.39(39) & 4.07(4) & 1.57(75)  & 6.9323 & 0.5822(9) & 1.827(6) & 2.30 & 3.973 & 1.574  & 3.33(11) & 35 & g20\\
 5090937 & 8.21(22) & 3.83(3) & 1.49(50) & 8.8007 & 0.241(13) & 6.50(19) & 2.10 & 3.909 & 1.455 & 6.25(20) & 22 & s16 \\
 5877364 & 7.39(16) & 3.94(3) & 1.18(8) & 89.649 & 0.8875(31) & 2.99(13) & 1.80 & 3.870 & 1.171 & 3.33(5) & 36 & s16 \\
 5960989 & 6.48(13) & 3.78(3) & 1.10(6)  & 50.722 & 0.810(16) & 3.80(50) & 1.70 & 3.812 & 1.086  & 4.54(22) & 31 & s16,d17\\
 6117415 & 6.40(13) & 4.04(3) & 0.70(48) & 19.742 & 0.7343(4) & 2.502(6) & 1.45 & 3.819 & 0.777 & 2.33(6) & 41 & c20 \\
 8027591 & 6.10(13) & 3.85(6) & 0.79(40) & 24.274 & 0.586(8) & 6.30(20) & 1.50 & 3.800 & 0.859 & 6.25(20) & 18 & s16 \\
 8164262 & 7.45(16) & 4.16(2) & 0.98(11) & 87.457 & 0.886(3) & 2.98(12) & 1.70 & 3.837 & 1.075  & 2.985 & 41 & h18 \\	
 8719324 & 7.28(17) & 3.96(3) & 1.13(17) & 10.233 & 0.5998(5) & 2.512(1) & 1.80 & 3.867 & 1.173 & 2.63(4) & 46 & g20\\
 9790355 & 6.42(13) & 3.79(4) & 1.06(12) & 14.566 & 0.513(7) & 4.96(12) & 1.70 & 3.812 & 1.086 & 4.54(22) & 31 & t12 \\
 9899216 & 7.30(17) & 3.63(3) & 1.56(55) & 10.916 & 0.65(1) & 2.15(10) & 2.20 & 3.883 & 1.559 & 1.67(3) & 104 & t12,s15 \\
 10334122 & 6.26(11) & 3.79(3) & 0.99(10) & 37.953 & 0.534(60) & 12.0(26) & 1.65 & 3.809 & 1.032 & 5.88(35) & 23 & s16 \\
11071278 & 5.90(12) & 3.72(4) & 0.90(19) & 55.885 & 0.755(14) & 6.23(57) & 1.60 & 3.803 & 0.980 & 6.67(23) & 19 & s16 \\
11403032 & 6.64(14) & 3.73(4) & 1.21(32) & 7.6316 & 0.288(13) & 5.05(16) & 1.80 & 3.822 & 1.193 & 10.5(6) & 14 & s16 \\  
11568657 & 6.37(12) & 3.84(4) & 0.97(16) & 13.476 & 0.565(2) & 3.80(3) & 1.60 & 3.809 & 0.974 & 5.13(13) & 24 & t12 \\  
11649962 & 6.80(15) & 4.06(3) & 0.81(33) & 10.563 & 0.5206(35) & 3.51(4)  & 1.50 & 3.831 & 0.830  &  3.03(10) & 32 & s16\\
11923629 & 6.23(13) & 4.11(4) & 0.56(62) & 17.973 & 0.3629(59) & 9.77(16) & 1.50 & 3.801 & 0.858 & 10.0(10) & 11 & s16 \\
12255108 & 7.72(18) & 4.05(3) & 1.20(18) & 9.1315 & 0.296(16) & 5.92(24) & 1.90 & 3.901 & 1.253 & 5.88(18)& 19 & s16 \\
		\hline
\multicolumn{13}{l}{$^{\rm a}$ Sources of eccentricity: t12 = \citet{Thompson2012}, ~s15 = \citet{Smullen2015}, ~s16 = \citet{Shporer2016},}\\ 
\multicolumn{13}{l}{\,d17 = \citet{Dimitrov2017}, ~g17 = \citet{Guo2017}, ~h18 = \citet{Hambleton2018}, ~g20 = \citet{Guo2020}, ~c20 = \citet{Cheng2020}} \\
\end{tabular}
\end{table*}

\begin{figure}
\includegraphics[width=0.49\textwidth]{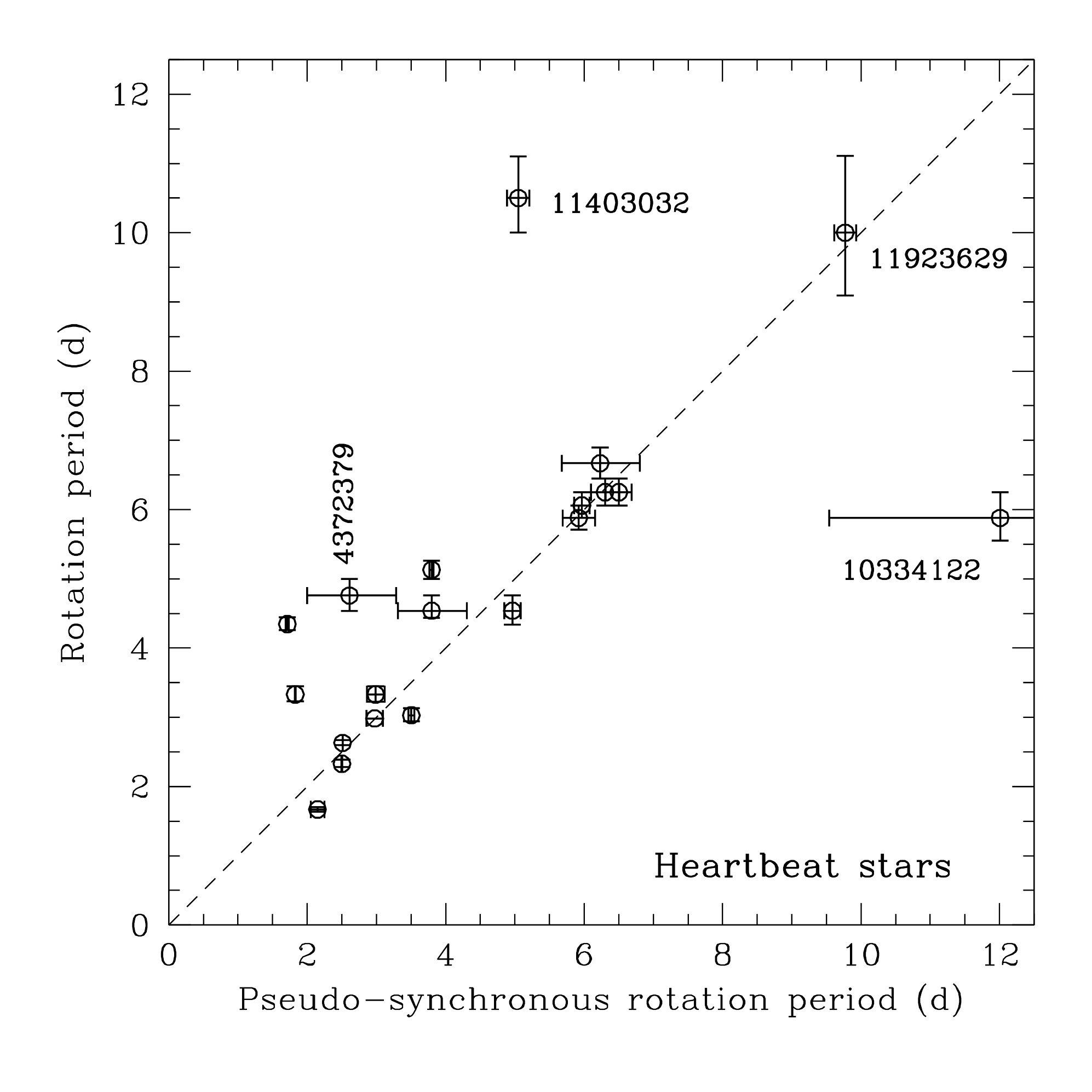}
\caption{Rotation period obtained by r-mode fitting versus pseudo-synchronous rotation period. }
\label{fig:rot}
\end{figure}

The results for each heartbeat system and the adopted model parameters are summarised in Table~\ref{tab:sum}. The obtained rotation periods are compared with the pseudo-synchronous rotation periods in Fig.\,\ref{fig:rot}. This figure shows that for the majority of the heartbeat systems in our samples, the rotation periods of primary stars are comparable to the corresponding pseudo-synchronous rotation, confirming the theory developed by  \citet{Hut1981}. (The rotation periods of KIC\,11403032 and KIC\,4372379 are comparable to the corresponding orbital period (see Fig.\,\ref{fig:rot_Porb} below)).

Our result is, however, different from the result of \citet{Zimmerman2017}, which concluded that the rotation periods tend to be approximately $\frac{3}{2}$ times the corresponding pseudo-synchronous rotation periods. The difference comes from \cite{Zimmerman2017} assuming that the centre of the main frequency group in the Fourier spectrum of a star is the rotation frequency, while we fit the group with $(m,k)=(1,-2)$ r~modes assuming a rotation frequency larger than the frequency of the main frequency group. Thus the rotation frequency (or period) obtained by \cite{Zimmerman2017} is systematically lower (or longer) than our value which is, in most cases, consistent with pseudo-synchronous rotation.  More specifically, as discussed in \S\ref{sec:freqrange}, the frequency range of the $(m,k)=(1,-2)$ r~modes in the observer's frame is between ${2\over3}f_{\rm rot}$ and $f_{\rm rot}$; or the period range is between ${3\over2}P_{\rm rot}$ and $P_{\rm rot}$.

\begin{figure*}
\includegraphics[width=0.49\textwidth]{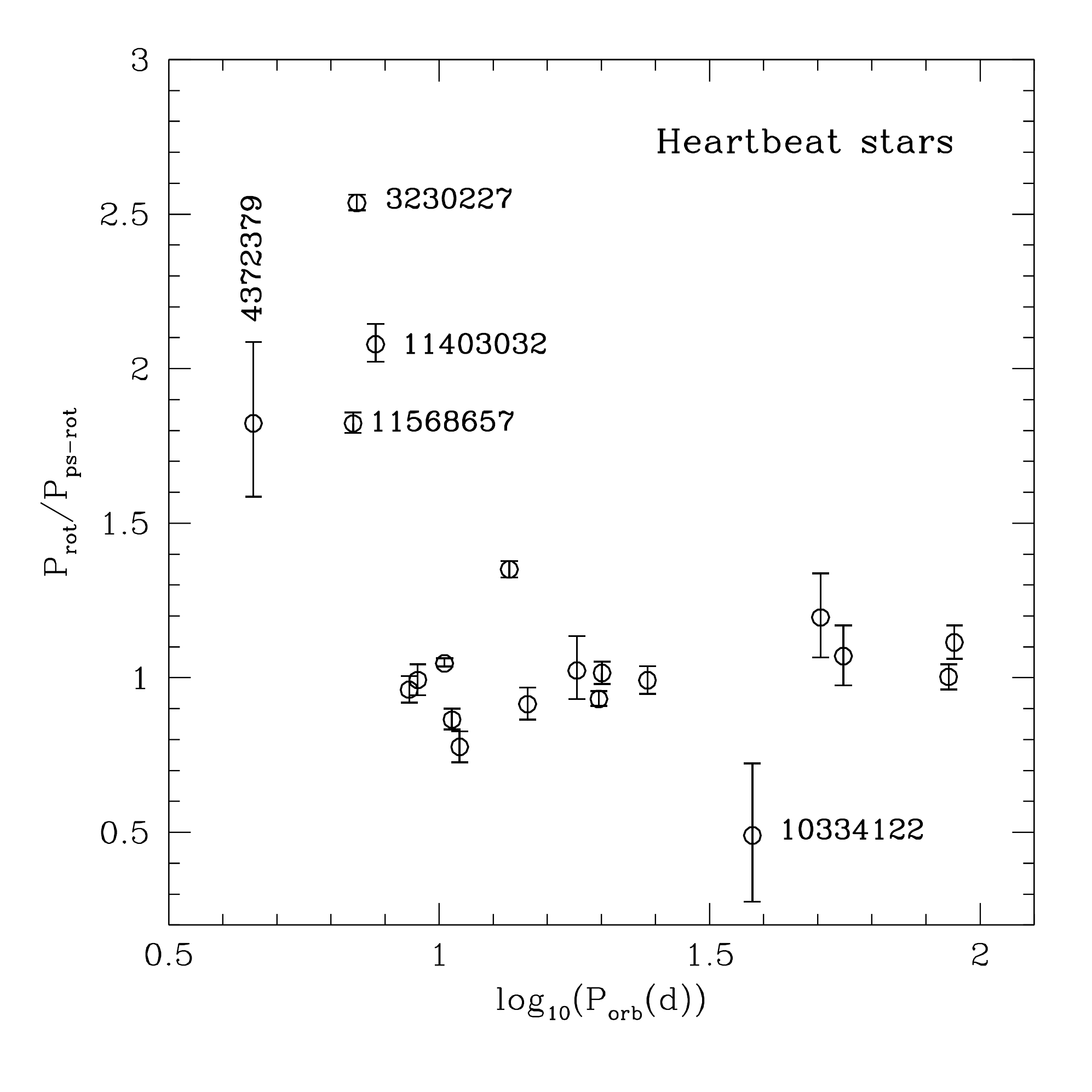} 
\includegraphics[width=0.49\textwidth]{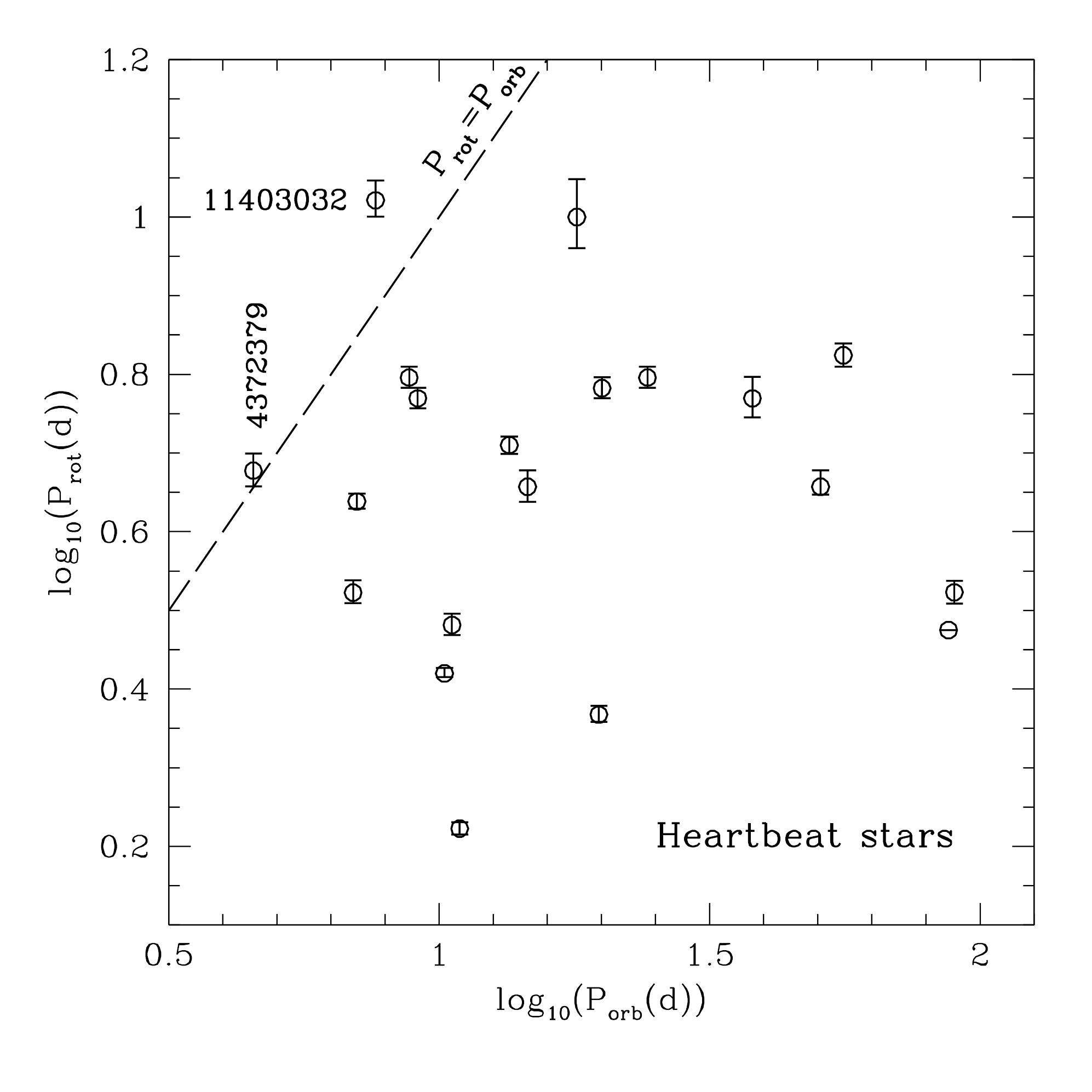}  
\caption{The left-hand panel shows the ratio of rotation period to the pseudo-synchronous rotation period versus the orbital period, while the right-hand panel shows the rotation period versus the  orbital period. }
\label{fig:rot_Porb}
\end{figure*}

The left panel of Fig.\,\ref{fig:rot_Porb} shows that in the heartbeat stars with orbital periods longer than 8\,d pseudo-synchronisation is roughly attained in all but  KIC\,10334122 (with large error bars), while all four systems of our sample with $P_{\rm orb}\la 8$\,d have rotation periods longer than the corresponding $P_{\rm ps-rot}$. Among the four systems with shorter orbital periods,  KIC\,4372379 and KIC\,11403032, which have small eccentricities of $0.342$ and $0.288$, respectively, have rotation periods comparable to the orbital periods (see right panel of Fig.\,\ref{fig:rot_Porb}); in particular, the rotation of KIC\,4372379 is almost synchronous with the orbital motion. The rotation periods of the other heartbeat stars in this figure are smaller than the corresponding orbital periods, which is because the angular velocity of the pseudo-synchronous rotation is comparable to the orbital angular velocity at the periastron \citep{Hut1981}. It is not clear to us why pseudo-synchronisation does not occur in the relatively short-period systems with the orbital periods shorter than 8\,d. 

\section{Conclusions}

Among the heartbeat stars observed by the {\it Kepler} satellite and collected by \citet{KEBC2016}, we have selected 20 stars whose FTs show densely gathered groups of frequencies that are consistent with the visibility distribution of r~modes, which have frequency ranges that depend on the rotation rate. For each star, we have obtained a rotation frequency by fitting the visibility distribution of $(m,k)=(1,-2)$ r~modes in a model whose parameters are consistent  with these given by \citet{Berger2020}. The rotation period thus obtained is, in most cases, found to be consistent  with the pseudo-synchronous rotation supporting the orbital evolution theory developed by \citet{Hut1981}. Scrutinising in more detail, we found that the rotation periods of all but one (KIC\,10334122 with large error bars) heartbeat binaries in our samples with $P_{\rm orb} > 8$\,d are close to the pseudo-synchronous rotation period. Conversely, all four heartbeat stars with $P_{\rm orb} < 8$\,d have rotation periods considerably larger than the pseudo-synchronous period. It is not clear at present the reason why the critical orbital period occurs at 8\,d for the pseudo-synchronisation of heartbeat stars. This is a particularly interesting problem, given that the rotation in these shorter orbital period systems are sub-synchronous, so angular momentum transfer has to continue to take rotational angular momentum and transfer it to the orbit after the system reaches pseudo-synchronous rotation.


\section*{Data Availability}
The data underlying this article will be shared on reasonable request to the corresponding author.
 


\bibliographystyle{mnras}
\bibliography{rotHBstar} 

\begin{thebibliography}{}
\makeatletter
\relax
\def\mn@urlcharsother{\let\do\@makeother \do\$\do\&\do\#\do\^\do\_\do\%\do\~}
\def\mn@doi{\begingroup\mn@urlcharsother \@ifnextchar [ {\mn@doi@}
  {\mn@doi@[]}}
\def\mn@doi@[#1]#2{\def\@tempa{#1}\ifx\@tempa\@empty \href
  {http://dx.doi.org/#2} {doi:#2}\else \href {http://dx.doi.org/#2} {#1}\fi
  \endgroup}
\def\mn@eprint#1#2{\mn@eprint@#1:#2::\@nil}
\def\mn@eprint@arXiv#1{\href {http://arxiv.org/abs/#1} {{\tt arXiv:#1}}}
\def\mn@eprint@dblp#1{\href {http://dblp.uni-trier.de/rec/bibtex/#1.xml}
  {dblp:#1}}
\def\mn@eprint@#1:#2:#3:#4\@nil{\def\@tempa {#1}\def\@tempb {#2}\def\@tempc
  {#3}\ifx \@tempc \@empty \let \@tempc \@tempb \let \@tempb \@tempa \fi \ifx
  \@tempb \@empty \def\@tempb {arXiv}\fi \@ifundefined
  {mn@eprint@\@tempb}{\@tempb:\@tempc}{\expandafter \expandafter \csname
  mn@eprint@\@tempb\endcsname \expandafter{\@tempc}}}

\bibitem[\protect\citeauthoryear{{Berger}, {Huber}, {van Saders}, {Gaidos},
  {Tayar}  \& {Kraus}}{{Berger} et~al.}{2020}]{Berger2020}
{Berger} T.~A.,  {Huber} D.,  {van Saders} J.~L.,  {Gaidos} E.,  {Tayar} J.,
  {Kraus} A.~L.,  2020, \mn@doi [\aj] {10.3847/1538-3881/159/6/280}, \href
  {https://ui.adsabs.harvard.edu/abs/2020AJ....159..280B} {159, 280}

\bibitem[\protect\citeauthoryear{{Cheng}, {Fuller}, {Guo}, {Lehman}  \&
  {Hambleton}}{{Cheng} et~al.}{2020}]{Cheng2020}
{Cheng} S.~J.,  {Fuller} J.,  {Guo} Z.,  {Lehman} H.,   {Hambleton} K.,  2020,
  \mn@doi [\apj] {10.3847/1538-4357/abb46d}, \href
  {https://ui.adsabs.harvard.edu/abs/2020ApJ...903..122C} {903, 122}

\bibitem[\protect\citeauthoryear{{Dimitrov}, {Kjurkchieva}  \&
  {Iliev}}{{Dimitrov} et~al.}{2017}]{Dimitrov2017}
{Dimitrov} D.~P.,  {Kjurkchieva} D.~P.,   {Iliev} I.~K.,  2017, \mn@doi
  [\mnras] {10.1093/mnras/stx745}, \href
  {https://ui.adsabs.harvard.edu/abs/2017MNRAS.469.2089D} {469, 2089}

\bibitem[\protect\citeauthoryear{{Guo}, {Gies}  \& {Fuller}}{{Guo}
  et~al.}{2017}]{Guo2017}
{Guo} Z.,  {Gies} D.~R.,   {Fuller} J.,  2017, \mn@doi [\apj]
  {10.3847/1538-4357/834/1/59}, \href
  {https://ui.adsabs.harvard.edu/abs/2017ApJ...834...59G} {834, 59}

\bibitem[\protect\citeauthoryear{{Guo}, {Shporer}, {Hambleton}  \&
  {Isaacson}}{{Guo} et~al.}{2020}]{Guo2020}
{Guo} Z.,  {Shporer} A.,  {Hambleton} K.,   {Isaacson} H.,  2020, \mn@doi
  [\apj] {10.3847/1538-4357/ab58c2}, \href
  {https://ui.adsabs.harvard.edu/abs/2020ApJ...888...95G} {888, 95}

\bibitem[\protect\citeauthoryear{{Hambleton} et~al.,}{{Hambleton}
  et~al.}{2018}]{Hambleton2018}
{Hambleton} K.,  et~al., 2018, \mn@doi [\mnras] {10.1093/mnras/stx2673}, \href
  {https://ui.adsabs.harvard.edu/abs/2018MNRAS.473.5165H} {473, 5165}

\bibitem[\protect\citeauthoryear{{Hut}}{{Hut}}{1981}]{Hut1981}
{Hut} P.,  1981, \aap, \href
  {https://ui.adsabs.harvard.edu/abs/1981A&A....99..126H} {99, 126}

\bibitem[\protect\citeauthoryear{{IJspeert}, {Tkachenko}, {Johnston}, {Garcia},
  {De Ridder}, {Van Reeth}  \& {Aerts}}{{IJspeert} et~al.}{2021}]{IJspeert2021}
{IJspeert} L.~W.,  {Tkachenko} A.,  {Johnston} C.,  {Garcia} S.,  {De Ridder}
  J.,  {Van Reeth} T.,   {Aerts} C.,  2021, \mn@doi [\aap]
  {10.1051/0004-6361/202141489}, \href
  {https://ui.adsabs.harvard.edu/abs/2021A&A...652A.120I} {652, A120}

\bibitem[\protect\citeauthoryear{{Kirk} et~al.,}{{Kirk}
  et~al.}{2016}]{KEBC2016}
{Kirk} B.,  et~al., 2016, \mn@doi [\aj] {10.3847/0004-6256/151/3/68}, \href
  {https://ui.adsabs.harvard.edu/abs/2016AJ....151...68K} {151, 68}

\bibitem[\protect\citeauthoryear{{Ko{\l}aczek-Szyma{\'n}ski}, {Pigulski},
  {Michalska}, {Mo{\'z}dzierski}  \&
  {R{\'o}{\.z}a{\'n}ski}}{{Ko{\l}aczek-Szyma{\'n}ski}
  et~al.}{2021}]{Kolaczek2021}
{Ko{\l}aczek-Szyma{\'n}ski} P.~A.,  {Pigulski} A.,  {Michalska} G.,
  {Mo{\'z}dzierski} D.,   {R{\'o}{\.z}a{\'n}ski} T.,  2021, \mn@doi [\aap]
  {10.1051/0004-6361/202039553}, \href
  {https://ui.adsabs.harvard.edu/abs/2021A&A...647A..12K} {647, A12}

\bibitem[\protect\citeauthoryear{{Kurtz}, {Shibahashi}, {Murphy}, {Bedding}  \&
  {Bowman}}{{Kurtz} et~al.}{2015}]{2015MNRAS.450.3015K}
{Kurtz} D.~W.,  {Shibahashi} H.,  {Murphy} S.~J.,  {Bedding} T.~R.,   {Bowman}
  D.~M.,  2015, \mn@doi [\mnras] {10.1093/mnras/stv868}, \href
  {https://ui.adsabs.harvard.edu/abs/2015MNRAS.450.3015K} {450, 3015}

\bibitem[\protect\citeauthoryear{{Lee}}{{Lee}}{2021}]{Lee2021}
{Lee} U.,  2021, \mn@doi [\mnras] {10.1093/mnras/stab1433}, \href
  {https://ui.adsabs.harvard.edu/abs/2021MNRAS.505.1495L} {505, 1495}

\bibitem[\protect\citeauthoryear{{Lee} \& {Saio}}{{Lee} \&
  {Saio}}{2020}]{Lee2020}
{Lee} U.,  {Saio} H.,  2020, \mn@doi [\mnras] {10.1093/mnras/staa2250}, \href
  {https://ui.adsabs.harvard.edu/abs/2020MNRAS.497.4117L} {497, 4117}

\bibitem[\protect\citeauthoryear{{Lenz} \& {Breger}}{{Lenz} \&
  {Breger}}{2005}]{Period04}
{Lenz} P.,  {Breger} M.,  2005, \mn@doi [Communications in Asteroseismology]
  {10.1553/cia146s53}, \href
  {https://ui.adsabs.harvard.edu/abs/2005CoAst.146...53L} {146, 53}

\bibitem[\protect\citeauthoryear{{Li}, {Van Reeth}, {Bedding}, {Murphy}  \&
  {Antoci}}{{Li} et~al.}{2019}]{LiG2019}
{Li} G.,  {Van Reeth} T.,  {Bedding} T.~R.,  {Murphy} S.~J.,   {Antoci} V.,
  2019, \mn@doi [\mnras] {10.1093/mnras/stz1171}, \href
  {https://ui.adsabs.harvard.edu/abs/2019MNRAS.487..782L} {487, 782}

\bibitem[\protect\citeauthoryear{{Paxton}, {Bildsten}, {Dotter}, {Herwig},
  {Lesaffre}  \& {Timmes}}{{Paxton} et~al.}{2011}]{pax11}
{Paxton} B.,  {Bildsten} L.,  {Dotter} A.,  {Herwig} F.,  {Lesaffre} P.,
  {Timmes} F.,  2011, \mn@doi [\apjs] {10.1088/0067-0049/192/1/3}, \href
  {https://ui.adsabs.harvard.edu/abs/2011ApJS..192....3P} {192, 3}

\bibitem[\protect\citeauthoryear{{Paxton} et~al.,}{{Paxton}
  et~al.}{2013}]{pax13}
{Paxton} B.,  et~al., 2013, \mn@doi [\apjs] {10.1088/0067-0049/208/1/4}, \href
  {http://adsabs.harvard.edu/abs/2013ApJS..208....4P} {208, 4}

\bibitem[\protect\citeauthoryear{{Paxton} et~al.,}{{Paxton}
  et~al.}{2015}]{pax15}
{Paxton} B.,  et~al., 2015, \mn@doi [\apjs] {10.1088/0067-0049/220/1/15}, \href
  {https://ui.adsabs.harvard.edu/abs/2015ApJS..220...15P} {220, 15}

\bibitem[\protect\citeauthoryear{{Saio}}{{Saio}}{2018}]{Saio2018}
{Saio} H.,  2018, arXiv e-prints, \href
  {https://ui.adsabs.harvard.edu/abs/2018arXiv181201253S} {p. arXiv:1812.01253}

\bibitem[\protect\citeauthoryear{{Saio}}{{Saio}}{2019}]{Saio2019}
{Saio} H.,  2019, \mn@doi [\mnras] {10.1093/mnras/stz1407}, \href
  {https://ui.adsabs.harvard.edu/abs/2019MNRAS.487.2177S} {487, 2177}

\bibitem[\protect\citeauthoryear{{Saio}}{{Saio}}{2020}]{Saio2020}
{Saio} H.,  2020, in {Neiner} C.,  {Weiss} W.~W.,  {Baade} D.,  {Griffin}
  R.~E.,  {Lovekin} C.~C.,   {Moffat} A.~F.~J.,  eds, Stars and their
  Variability Observed from Space. pp 321--324 (\mn@eprint {arXiv}
  {1912.00705})

\bibitem[\protect\citeauthoryear{{Saio}, {Kurtz}, {Murphy}, {Antoci}  \&
  {Lee}}{{Saio} et~al.}{2018}]{Saio+2018}
{Saio} H.,  {Kurtz} D.~W.,  {Murphy} S.~J.,  {Antoci} V.~L.,   {Lee} U.,  2018,
  \mn@doi [\mnras] {10.1093/mnras/stx2962}, \href
  {https://ui.adsabs.harvard.edu/abs/2018MNRAS.474.2774S} {474, 2774}

\bibitem[\protect\citeauthoryear{{Saio}, {Takata}, {Lee}, {Li}  \& {Van
  Reeth}}{{Saio} et~al.}{2021}]{Saio2021}
{Saio} H.,  {Takata} M.,  {Lee} U.,  {Li} G.,   {Van Reeth} T.,  2021, \mn@doi
  [\mnras] {10.1093/mnras/stab482}, \href
  {https://ui.adsabs.harvard.edu/abs/2021MNRAS.502.5856S} {502, 5856}

\bibitem[\protect\citeauthoryear{{Shporer} et~al.,}{{Shporer}
  et~al.}{2016}]{Shporer2016}
{Shporer} A.,  et~al., 2016, \mn@doi [\apj] {10.3847/0004-637X/829/1/34}, \href
  {https://ui.adsabs.harvard.edu/abs/2016ApJ...829...34S} {829, 34}

\bibitem[\protect\citeauthoryear{{Smullen} \& {Kobulnicky}}{{Smullen} \&
  {Kobulnicky}}{2015}]{Smullen2015}
{Smullen} R.~A.,  {Kobulnicky} H.~A.,  2015, \mn@doi [\apj]
  {10.1088/0004-637X/808/2/166}, \href
  {https://ui.adsabs.harvard.edu/abs/2015ApJ...808..166S} {808, 166}

\bibitem[\protect\citeauthoryear{{Thompson} et~al.,}{{Thompson}
  et~al.}{2012}]{Thompson2012}
{Thompson} S.~E.,  et~al., 2012, \mn@doi [\apj] {10.1088/0004-637X/753/1/86},
  \href {https://ui.adsabs.harvard.edu/abs/2012ApJ...753...86T} {753, 86}

\bibitem[\protect\citeauthoryear{{Van Reeth}, {Tkachenko}  \& {Aerts}}{{Van
  Reeth} et~al.}{2016}]{VanReeth2016}
{Van Reeth} T.,  {Tkachenko} A.,   {Aerts} C.,  2016, \mn@doi [\aap]
  {10.1051/0004-6361/201628616}, \href
  {https://ui.adsabs.harvard.edu/abs/2016A&A...593A.120V} {593, A120}

\bibitem[\protect\citeauthoryear{{Wrona} et~al.,}{{Wrona}
  et~al.}{2021}]{Wrona2021}
{Wrona} M.,  et~al., 2021, arXiv e-prints, \href
  {https://ui.adsabs.harvard.edu/abs/2021arXiv210914616W} {p. arXiv:2109.14616}

\bibitem[\protect\citeauthoryear{{Zimmerman}, {Thompson}, {Mullally}, {Fuller},
  {Shporer}  \& {Hambleton}}{{Zimmerman} et~al.}{2017}]{Zimmerman2017}
{Zimmerman} M.~K.,  {Thompson} S.~E.,  {Mullally} F.,  {Fuller} J.,  {Shporer}
  A.,   {Hambleton} K.,  2017, \mn@doi [\apj] {10.3847/1538-4357/aa85e3}, \href
  {https://ui.adsabs.harvard.edu/abs/2017ApJ...846..147Z} {846, 147}

\makeatother
\end{thebibliography}



\appendix

\section{Other heartbeat stars fitted with r~modes}

\subsection{KIC\,4372379}\label{sec:k437}

\begin{figure}
\includegraphics[width=0.49\textwidth]{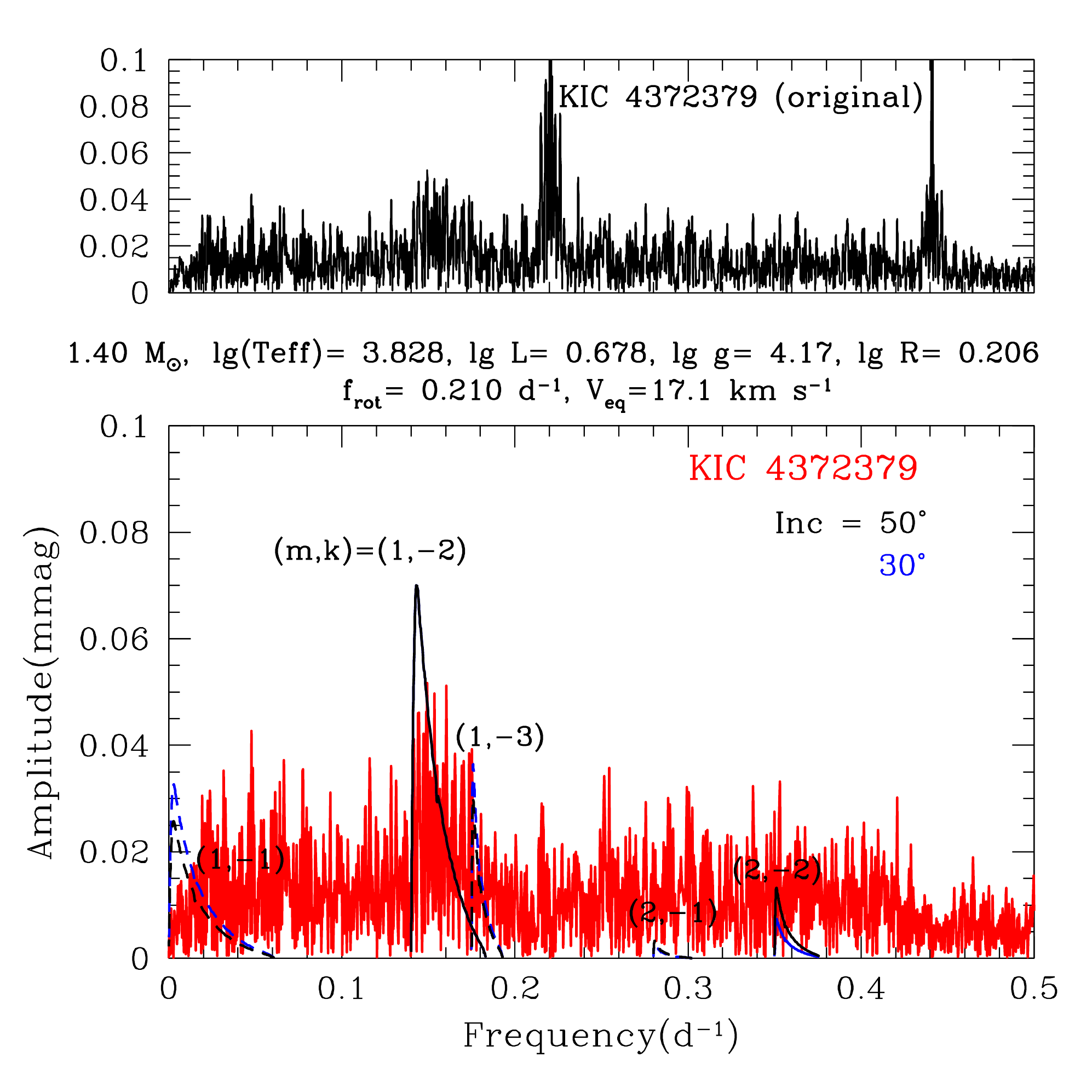}
\caption{FTs for KIC\,4372379 in the same format as Fig.\,\ref{fig:k816}. \citet{Thompson2012} obtained the orbital inclination to be $48 \pm 11^\circ$ for this star.} 
\label{fig:k437}
\end{figure}

KIC\,4372379 is a short period ($P_{\rm orb}= 4.535$\,d; KEBC) heartbeat binary with an orbital eccentricity $0.342 \pm 0.092$ \citep{Thompson2012}. From equation~(\ref{eq:ps}) we obtain $P_{\rm ps-rot}= 2.61\pm0.64$\,d. Fig.\,\ref{fig:k437} (lower panel) shows an FT for KIC\,4372379. Also shown are visibility distributions of r~modes predicted for a 1.40-M$_\odot$ model at a rotation frequency of $0.21$\,d$^{-1}$  ($P_{\rm rot}=4.76$\,d). The rotation frequency has been chosen to reproduce the frequency group around $0.14 - 0.18$\,d$^{-1}$ by the expected frequency ranges of $(m,k)=(1,-2)$ and $(1,-3)$ r~modes. The rotation period is longer than $P_{\rm ps-rot}$, while close to the orbital period (see Fig.\,\ref{fig:rot_Porb}).

\subsection{KIC\,5034333}\label{sec:k503}

\begin{figure}
\includegraphics[width=0.49\textwidth]{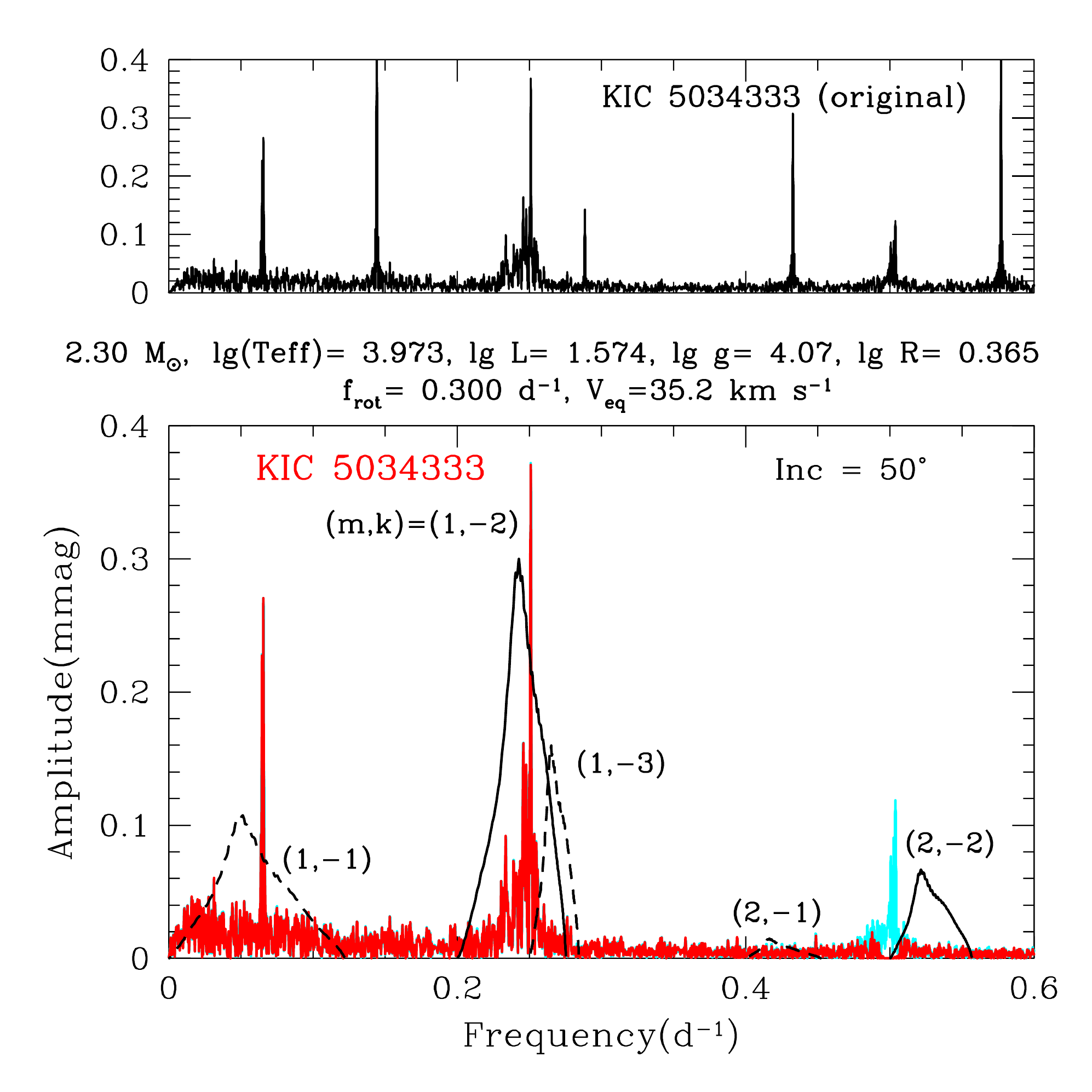}
\caption{FTs for KIC\,5034333 and r~mode visibility distributions for the indicated parameters. The cyan line denotes harmonics and combination frequencies of high-amplitude ($> 8$~mmag) peaks at $\sim$$0.25$\,d$^{-1}$.  We note that \citet{Guo2020} obtained an orbital inclination of $49.88^\circ$.}
\label{fig:k503}
\end{figure}

The orbital period of KIC\,5034333 is $6.932$\,d (KEBC), while \citet{Guo2020} obtained an orbital eccentricity of $0.5822 \pm 0.0009$.
\footnote{ 
While \citet{Thompson2012} obtained $e=0.575\pm0.002$ using a similar method applied to the {\it Kepler} data between quaters 0 and 7, we have adopted in Table\,\ref{tab:sum} the result of \citet{Guo2020} based  on a longer {\it Kepler} data set from quaters 0 to 17.}
Using these parameters in equation~(\ref{eq:ps}) yields $P_{\rm ps-rot}=1.827\pm0.006$\,d. Fig.\,\ref{fig:k503} (lower panel) shows an FT for KIC\,5034333 with r-mode visibility distributions predicted for a 2.30-M$_\odot$ model at a rotation frequency of $0.30$\,d$^{-1}$ ($P_{\rm rot}= 3.33$\,d). The rotation frequency is chosen to fit the frequency range of the group at $\sim$$0.25$\,d$^{-1}$ with r~modes of $(m,k)=(1,-2)$, while \citet{Zimmerman2017} obtained a slightly slower rotation frequency of $0.251$\,d$^{-1}$,  considering the strongest peak at that frequency to be caused by a spot on the stellar surface.

\subsection{KIC\,5090937}\label{sec:k509}

\begin{figure}
\includegraphics[width=0.49\textwidth]{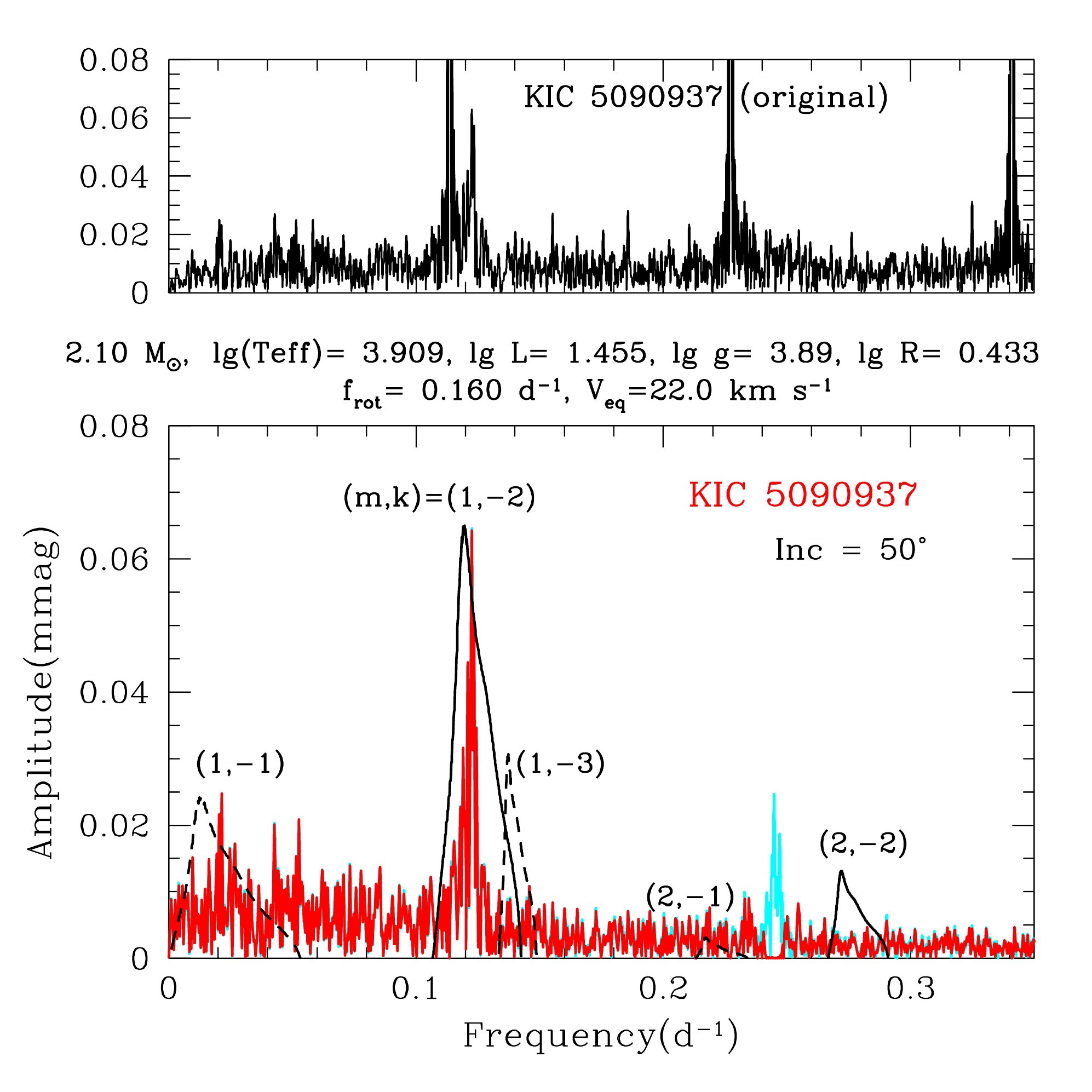}
\caption{FTs for KIC\,5090937. The cyan line denotes harmonics and combination frequencies of the relatively high amplitude ($>0.03$~mmag) peaks around $0.12$\,d$^{-1}$.}
\label{fig:k509}
\end{figure}

KIC\,5090937 is a heartbeat star with a relatively short orbital period of 8.8007\,d (KEBC) and a relatively small orbital eccentricity of  $0.241$ \citep{Shporer2016}. Using these orbital parameters in equation (\ref{eq:ps}) yields $P_{\rm ps-rot}=6.50$\,d. Fig.\,\ref{fig:k509} (lower panel) shows an FT for KIC\,5090937 with r-mode visibility distribution for a 2.10-M$_\odot$ model. The frequency group at $\sim$$0.12$\,d$^{-1}$ is fitted with r~modes of $(m,k)=(1,-2)$ at a rotation frequency of $0.160$\,d$^{-1}$. The determined rotation frequency corresponds to $P_{\rm rot}=6.25$\,d, which is comparable to $P_{\rm ps-rot}$.

\subsection{KIC\,5877364 (HD 183613)}\label{sec:k587}

\begin{figure}
\includegraphics[width=0.49\textwidth]{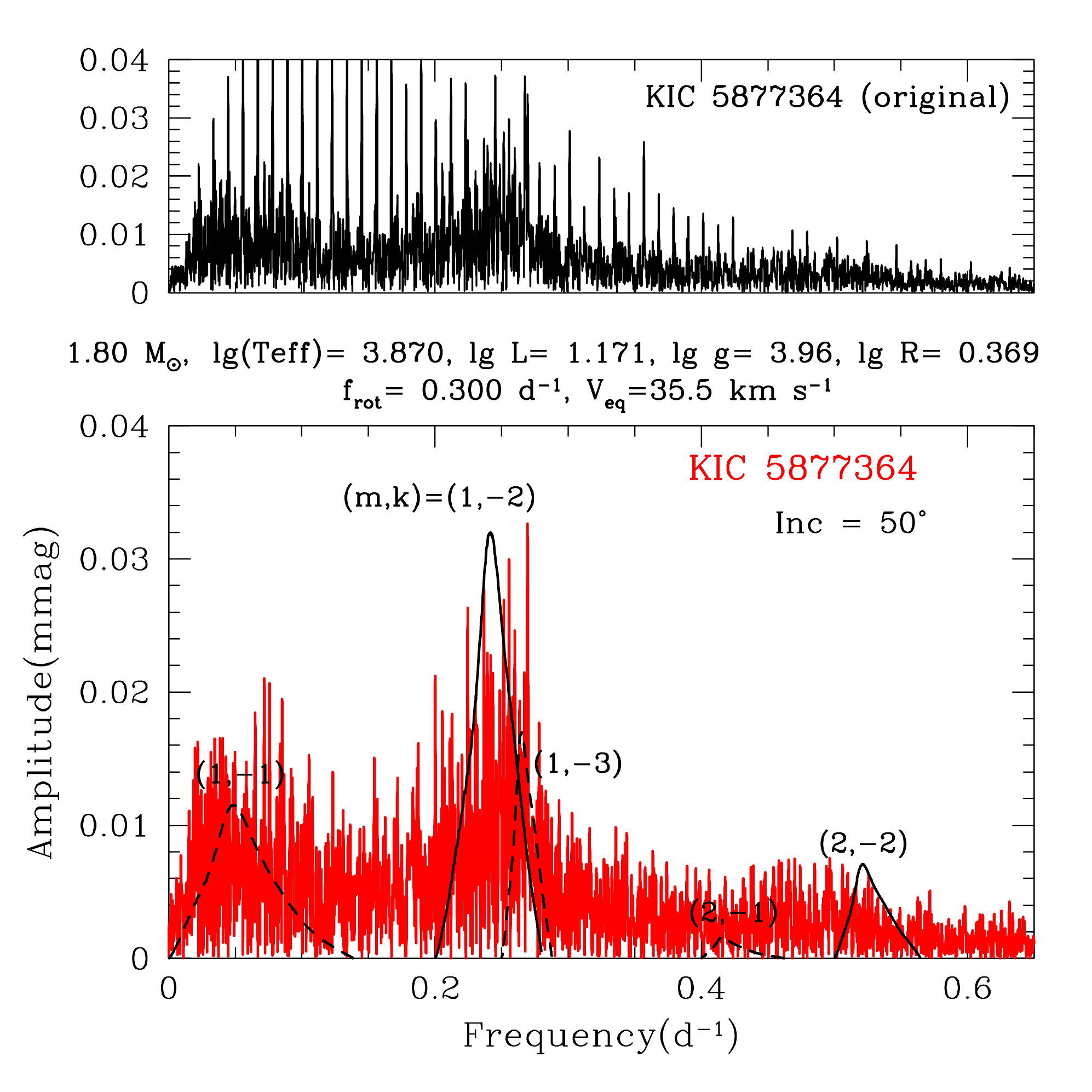}
\caption{FTs for KIC\,5877364 and  visibility distributions of r~modes in a 1.80-M$_\odot$ model at $f_{\rm rot}=0.30$\,d$^{-1}$.
}
\label{fig:k587}
\end{figure}

KIC\,5877364 has a long orbital period of $89.649$\,d (KEBC) ($f_{\rm orb}=0.0112$\,d$^{-1}$) with a very large orbital eccentricity of $e=0.8875\pm 0.031$ \citep{Shporer2016}, which yields the pseudo-synchronous period $P_{\rm ps-rot} = 2.99 \pm 0.13$\,d. Fig.\,\ref{fig:k587} shows FTs for KIC\,5877364. After removing dense orbital peaks (upper panel), a clear group of frequencies at $\sim$$0.2 - 0.3$\,d$^{-1}$ (red line) is apparent (lower panel), to which  we fit the frequency range of $(m,k)=(1,-2)$ r~modes in a 1.80-M$_\odot$ model at a rotation frequency of $0.30$\,d$^{-1}$ ($P_{\rm rot}= 3.33$\,d). The rotation period is consistent with the pseudo-synchronous rotation period. Because of the very long orbital period, many orbital harmonic frequencies lie within  r-mode frequency ranges so that the amplitudes of these r~modes  might be enhanced by tidal effects.
 
\subsection{KIC\,5960989}\label{sec:k596}

\begin{figure}
\includegraphics[width=0.49\textwidth]{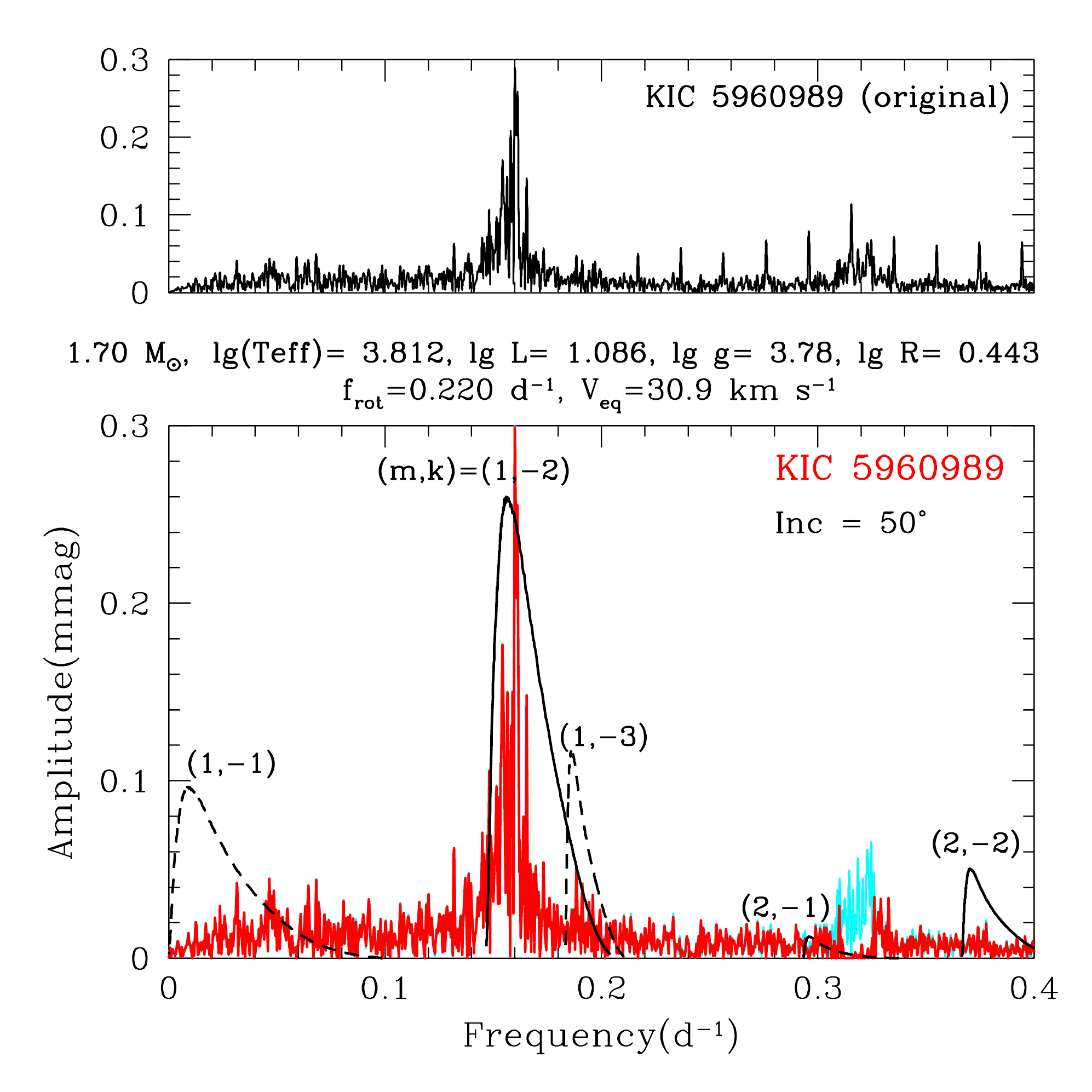}
\caption{FTs for  KIC\,5960989 and visibility distributions of r~modes in a 1.70-M$_\odot$ model at a rotation frequency of $0.22$\,d$^{-1}$. The cyan line denotes frequencies identified as harmonics and combination frequencies of the frequencies whose amplitudes  are larger than $0.1$~mmag. The case of inclination $50^\circ$ is shown because \citet{Dimitrov2017} obtained an orbital inclination of $51.0\pm0.3^{\circ}$.}
\label{fig:k596}
\end{figure}

The orbital period of the heartbeat binary KIC\,5960989 is $50.722$\,d (KEBC), while the eccentricity $0.810\pm0.016$ is adopted in Table~\ref{tab:sum} by avaraging the results of \cite{Shporer2016} and \citet{Dimitrov2017}. Using these orbital parameters in equation~(\ref{eq:ps}), we obtain $P_{\rm ps-rot}=3.80\pm0.50$\,d. Fig.\,\ref{fig:k596} (lower panel) shows an FT for  KIC\,5960989. Also shown are visibility distributions of r~modes in a 1.7-M$_\odot$ model (see Table~\ref{tab:sum}) at a rotation frequency, $f_{\rm rot}=0.22$\,d$^{-1}$, which is chosen for the frequency range of $(m,k)=(1,-2)$ r~modes to be consistent with the frequency group around $0.16$\,d$^{-1}$. The rotation frequency corresponds to the period $4.54$\,d, which is comparable to the pseudo-synchronous rotation period. 
  
\subsection{KIC\,6117415}\label{sec:k611}

\begin{figure}
\includegraphics[width=0.49\textwidth]{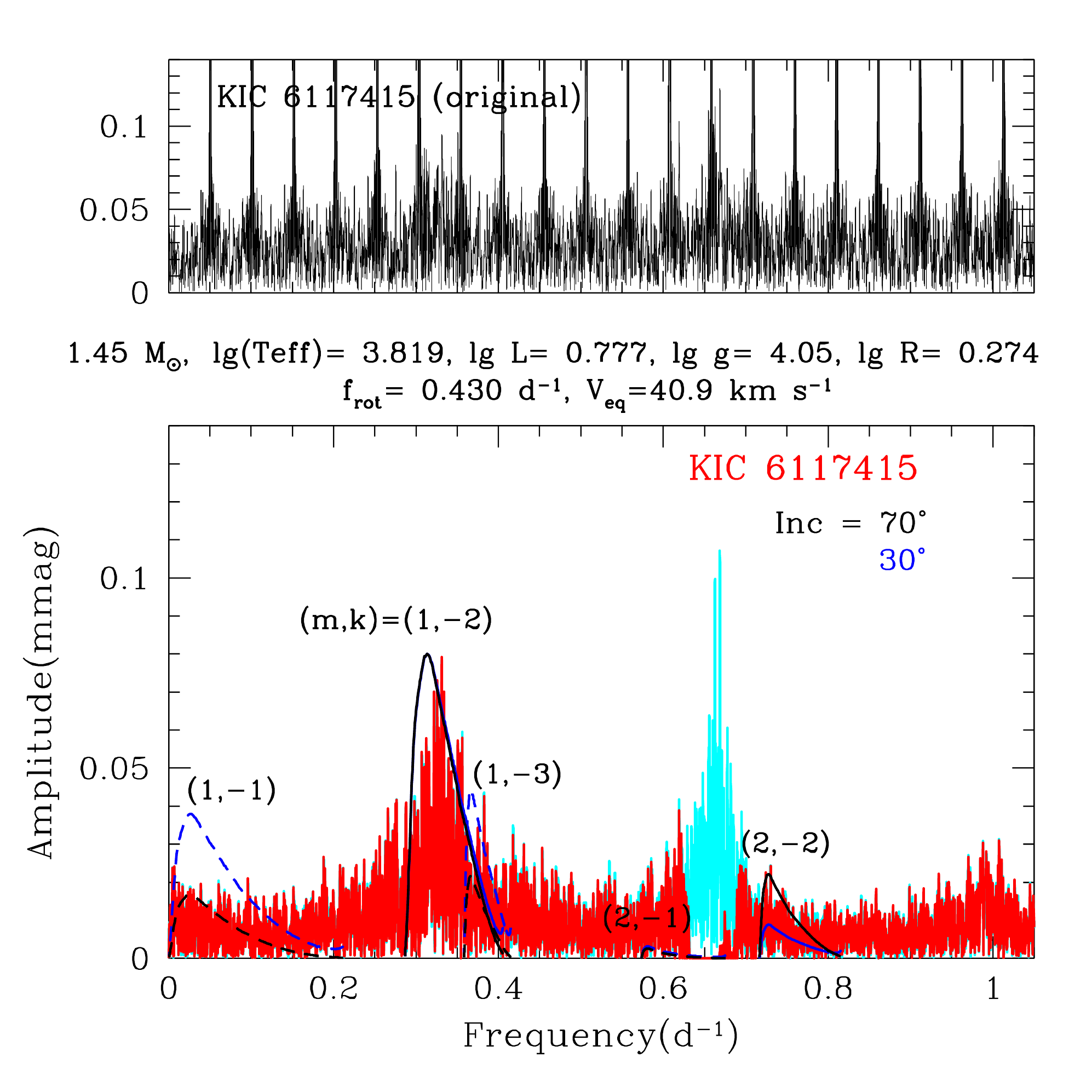}
\caption{FTs for  KIC\,6117415 and theoretical visibility distributions of a model at a rotation frequency of $0.43$\,d$^{-1}$.
The cyan line shows harmonics and combination frequencies of high-amplitude peaks ($\ga 0.06$~mmag) in the group at  $\sim$$0.3$\,d$^{-1}$. The visibility of r~modes is normalised as $0.08$~mmag at the maximum ($\sim$$0.3$\,d$^{-1}$) of $(m,k)=(1,-2)$ r~modes.  It can be seen that some of the combination frequencies have higher amplitudes than the parent peaks. \citet[][section 2.3]{2015MNRAS.450.3015K} discussed this and showed theoretically how it occurs.}
\label{fig:k611}
\end{figure}

KIC\,6117415 is a heartbeat system with an orbital period of $19.742$\,d (KEBC) ($f_{\rm orb}=0.0507$\,d$^{-1}$). Fig.\,\ref{fig:k611} shows FTs for KIC\,6117415 in a similar format to that of Fig.\,\ref{fig:k816}. R-mode visibility distributions are shown for a 1.45-M$_\odot$ main-sequence model at a rotation frequency of $0.43$\,d$^{-1}$ assuming rotational inclination angles of $70^\circ$ and $30^\circ$. The frequency group at $0.3 - 0.35$\,d$^{-1}$ is well fitted with $(m,k)=(1,-2)$ r~modes at $f_{\rm rot}=0.43$\,d$^{-1}$.

\citet{Cheng2020} obtained an orbital eccentricity of $0.734$ for KIC~6117415. Using that value of eccentricity and the orbital period $19.742$~d in equation (\ref{eq:ps}), we obtain $P_{\rm ps-rot}= 2.502$~d, which is comparable to the rotation period $2.33$~d that corresponds to the assumed rotation frequency $f_{\rm rot}=0.43$~d$^{-1}$. \citet{Cheng2020} also obtained an orbital inclination of $83.2^\circ$ and a projected rotation velocity of $v\sin i= 19 - 20$~km~s$^{-1}$ for KIC~6117415. Since our model predicts an equatorial rotation velocity of $41$~km~s$^{-1}$, the inclination of the rotational axis should be about $30^{\circ}$, which is inclined to the orbital axis by about $50^{\circ}$.  
(Such a misalignment is also found in KIC\,8164262 as discussed in \S\,\ref{sec:k816}.)  
For the rotational inclination of $30^\circ$ (blue line in Fig.~\ref{fig:k611}), amplitudes of $(m,k)=(1,-1)$ r modes near to 0\,d$^{-1}$ are predicted to be larger (under the energy equipartition assumption) than observed, which could indicate the antisymmetric r modes to be less strongly excited. \citet{Cheng2020} also analysed another two heartbeat stars, KIC~11494130 and KIC~5790807. However, no clear r-mode features are found in the FTs of those binaries.

\subsection{KIC\,8027591}\label{sec:k802}

\begin{figure}
\includegraphics[width=0.49\textwidth]{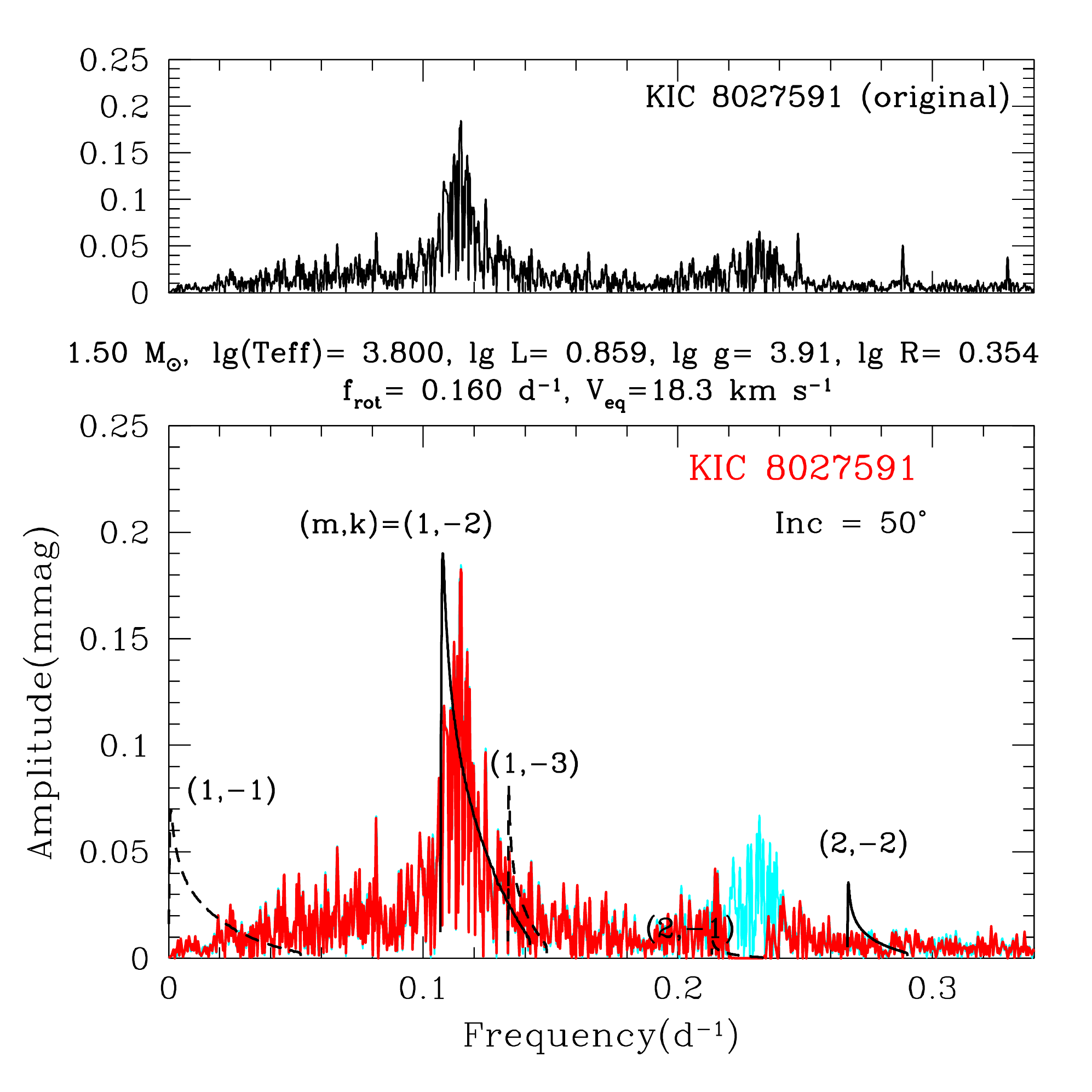}
\caption{FTs for KIC\,8027591. The cyan line in the lower panel denotes harmonics and combination frequencies resulting from frequencies in the main group with amplitudes larger than $0.1$~mmag. }
\label{fig:k802}
\end{figure}

The heartbeat binary KIC\,8027591 has an orbital period $24.274$\,d (KEBC), and \citet{Shporer2016} obtained an orbital eccentricity of $0.586\pm0.008$. From these parameters equation~(\ref{eq:ps}) yields $P_{\rm ps-rot}=6.30\pm0.20$\,d. Fig.\,\ref{fig:k802} (lower panel) shows an FT for KIC\,8027591 and predicted visibility distributions for r~modes in a 1.50-M$_\odot$ model at a rotation frequency of $0.16$\,d$^{-1}$ ($P_{\rm rot} = 6.25$\,d), which is chosen to make the frequency range of $(m,k)=(1,-2)$ r~modes consistent with the observed frequency group around $0.12$\,d$^{-1}$. The obtained rotation period agrees with the $P_{\rm ps-rot}$. 

\subsection{KIC\,8719324}\label{sec:k871}

\begin{figure}
\includegraphics[width=0.49\textwidth]{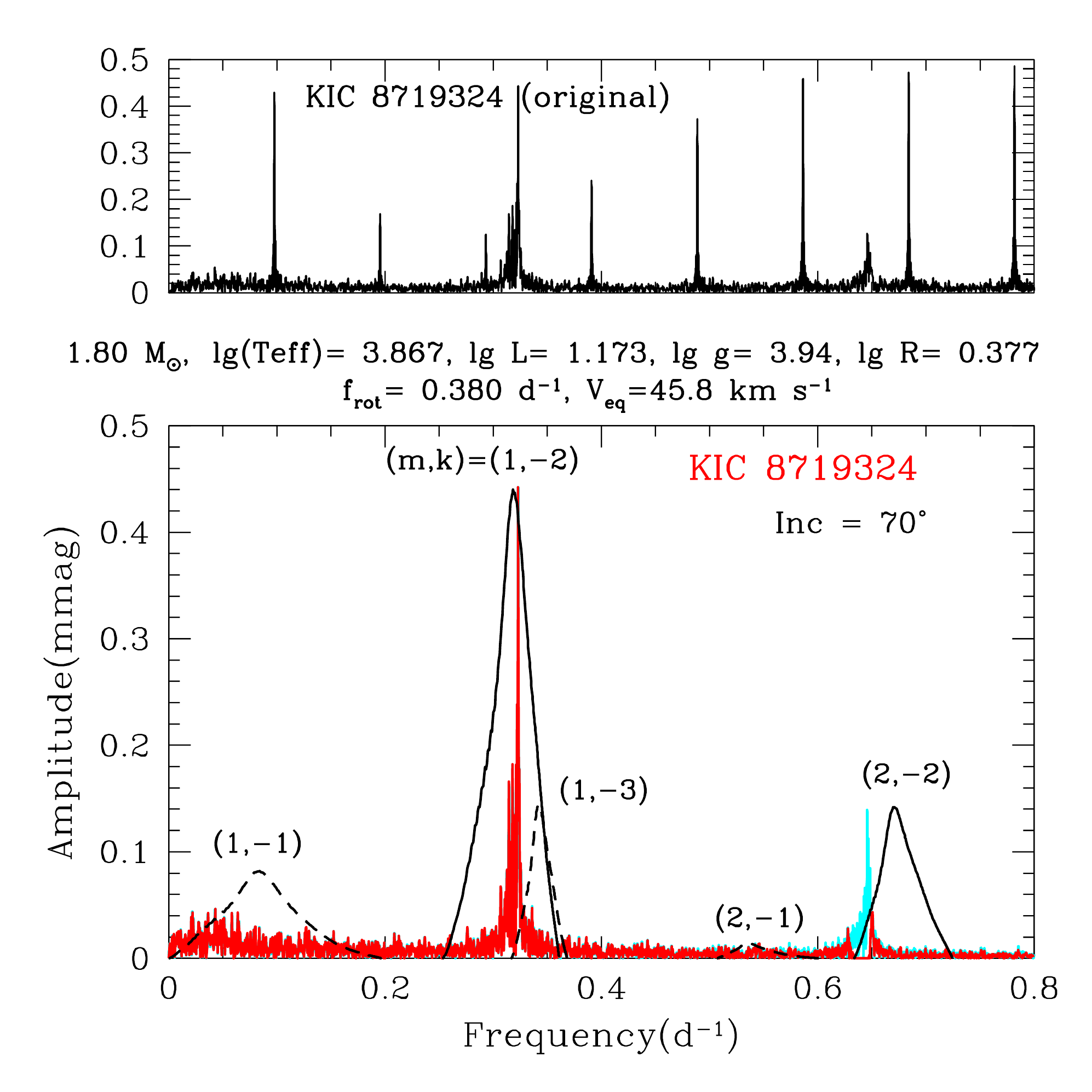}
\caption{FTs for  KIC\,8719324; the upper panel includes orbital harmonics, which are removed in the lower panel. The cyan line denotes harmonics and combination frequencies of the highest peaks ($\ga 0.1$~mmag) of the frequency group at $0.32$\,d$^{-1}$.  \citet{Guo2020} obtained an orbital inclination of $73.54\pm0.06^\circ$. They also found tidally excited pulsations at $\sim$$2.5$ and $\sim$$3$\,d$^{-1}$, which are out of the range of this diagram. }
\label{fig:k871}
\end{figure}

KIC\,8719324 is a heartbeat binary with an orbital period of $10.233$\,d (KEBC) and an eccentricity of $0.5998 \pm 0.0001$ \citep{Guo2020}
\footnote{  
The eccentricity was obtained by fixing $P_{\rm orb}$ at the KEBC value, while \citet{Smullen2015} obtained $e=0.64\pm0.05$ and $P_{\rm orb}=10.235\pm 0.005$\,d.  We have adopted, in Table\,\ref{tab:sum}, the eccentricity based on the fixed $P_{\rm orb}$.}
, which yield $P_{\rm ps-rot}= 2.512$\,d from equation~(\ref{eq:ps}).  Fig.\,\ref{fig:k871} shows FTs for KIC\,8719324; as in the other figures, the top panel shows the FT obtained from the original data including orbital harmonics, which are removed in the lower panel.
   
\citet{Zimmerman2017} identified  the peaks at $\sim$$0.323$ and $\sim$$0.646$\,d$^{-1}$ as rotational modulations,  hence obtained the rotation period of KIC\,8719324 to be $3.1$\,d. Since the above mentioned features are not single peaks but consist of  dense groups of peaks, we fit the group of peaks at $\sim$$0.323$\,d$^{-1}$ with $(m,k)=(1,-2)$ r~modes in a 1.80-M$_\odot$ model at a rotation frequency of $0.38$\,d$^{-1}$ ($P_{\rm rot}= 2.63$\,d). Our rotation period,  $2.63$\,d, is similar to the pseudo-synchronous rotation $2.51$\,d.

\subsection{KIC\,9790355}\label{sec:k979} 

\begin{figure}
\includegraphics[width=0.49\textwidth]{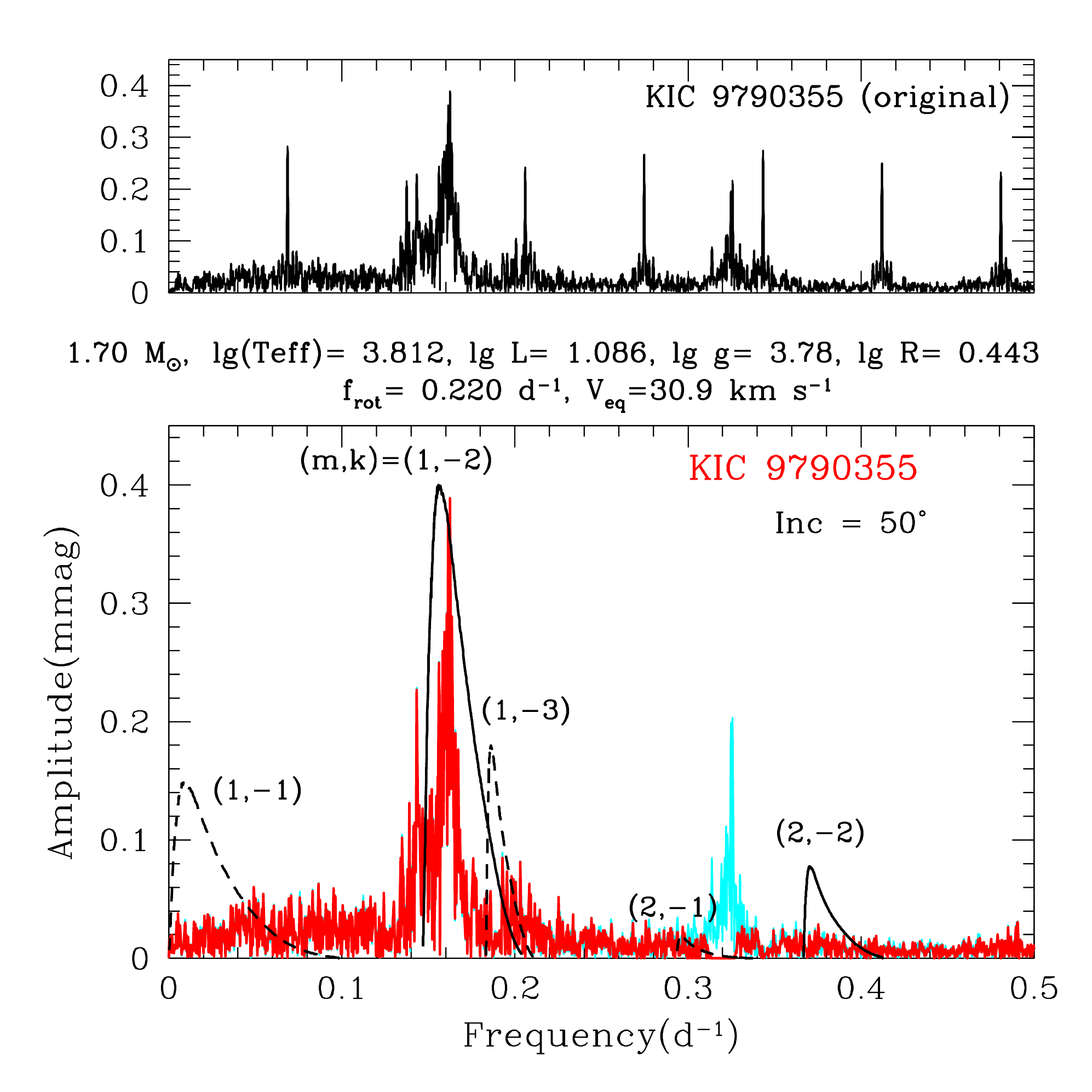}
\caption{FTs for KIC\,9790355. Harmonics and combination frequencies of high-amplitude ($>0.2$\,mmag) peaks at $\sim$$0.16$\,d$^{-1}$ are shown by the cyan line. R-mode visibilities for an inclination of $50^\circ$ are shown because \citet{Thompson2012} obtained an orbital inclination of $46.3\pm0.7^{\circ}$ for this star. }
\label{fig:k979}
\end{figure}

The orbit of the heartbeat binary KIC\,9790355 has a period of 14.566\,d (KEBC) and an eccentricity of $0.513\pm0.007$ \citep{Thompson2012}. From these parameters equation~(\ref{eq:ps}) gives a pseudo-synchronous period $P_{\rm ps-rot}= 4.96\pm 0.12$\,d. Fig.\,\ref{fig:k979} shows FTs for KIC\,9790355 with r-mode visibility distributions for a 1.80-M$_\odot$ model in a similar format to Fig.\,\ref{fig:k816}. In order to fit the frequency group at $\sim$$0.16$ with the frequency range of $(m,k)=(1,-2)$ r~modes, a rotation frequency of $0.22$\,d$^{-1}$ is adopted. The corresponding rotation period, $4.76$\,d, is similar to $P_{\rm ps-rot}$ of KIC\,9790355.

\subsection{KIC\,10334122}\label{sec:k103}

\begin{figure}
\includegraphics[width=0.49\textwidth]{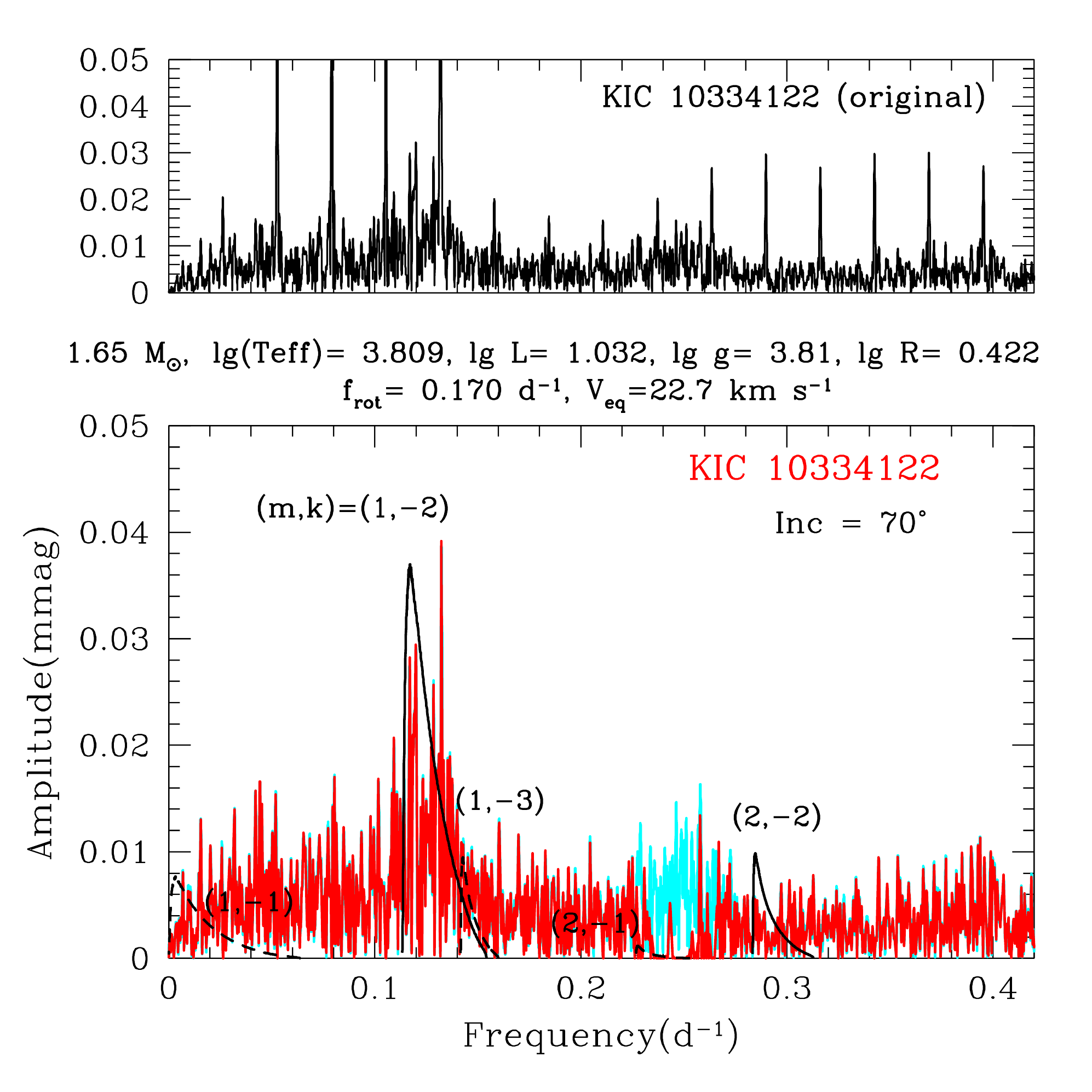}
\caption{FTs for KIC\,10334122. The cyan line indicates harmonics and combination frequencies of ten highest amplitude ($> 0.02$ mmag) peaks in the frequency group around 0.12\,d$^{-1}$, which is fitted with $(m,k)=(1,-2)$ at a rotation frequency of 0.17\,d$^{-1}$.}
\label{fig:103}
\end{figure}

KIC\,10334122 is a relatively long period ($P_{\rm orb}=37.953$\,d; KEBC) heartbeat star. The orbital eccentricity was determined by \citet{Shporer2016} to be $0.53 \pm 0.06$. Using the orbital period and eccentricity in equation~(\ref{eq:ps}) yields $P_{\rm ps-rot}= 12.0 \pm 2.6$\,d, which is considerably longer than the rotation period $5.88$\,d obtained by fitting the frequency group at $\sim$$0.12$\,d$^{-1}$ by the $(m,k)=(1,-2)$ r~modes (Fig.\,\ref{fig:103}). Among our sample heartbeat stars, this is the only case where $P_{\rm ps-rot} \gg P_{\rm rot}$. 

\subsection{KIC\,11071278}\label{sec:k110}

\begin{figure}
\includegraphics[width=0.49\textwidth]{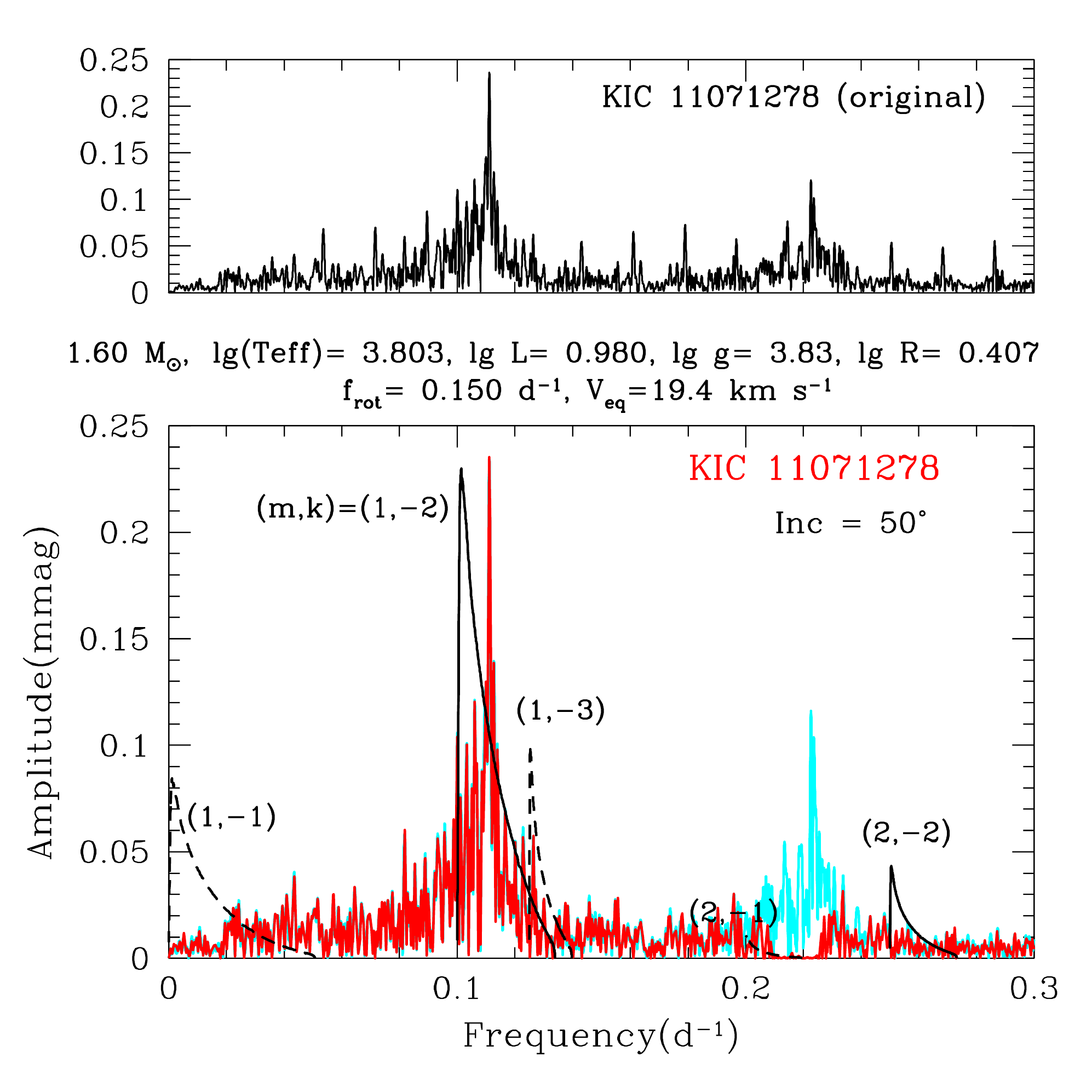}
\caption{FTs for  KIC\,11071278. The cyan line indicates harmonics and combination frequencies of the highest-amplitude peaks ($> 0.1$ mmag) around $0.11$\,d$^{-1}$.  }
\label{fig:k110}
\end{figure}

KIC\,11071278 is a relatively cool \citep[$T_{\rm eff} = 5900\pm120$\,K;][]{Berger2020} heartbeat binary with an orbital period of $55.885$\,d (KEBC). \citet{Shporer2016} estimated an orbital eccentricity of $0.755$. With these parameters equation~(\ref{eq:ps}) gives a pseudo-synchronous rotation period $P_{\rm ps-rot}= 6.23$\,d. Fig.\,\ref{fig:k110} (lower panel) shows an FT for  KIC\,11071278. Regarding all frequencies in the hump at $\sim$$0.22$\,d$^{-1}$ as harmonics and combination frequencies, we fit the main hump at $\sim$$0.11$\,d$^{-1}$ with $(m,k)=(1,-2)$ r~modes, assuming a rotation frequency of $0.15$\,d$^{-1}$ ($P_{\rm rot}= 6.67$\,d). The assumed rotation frequency is consistent with $P_{\rm ps-rot}$ for KIC\,11071278.

\subsection{KIC\,11403032}\label{sec:k114}

\begin{figure}
\includegraphics[width=0.49\textwidth]{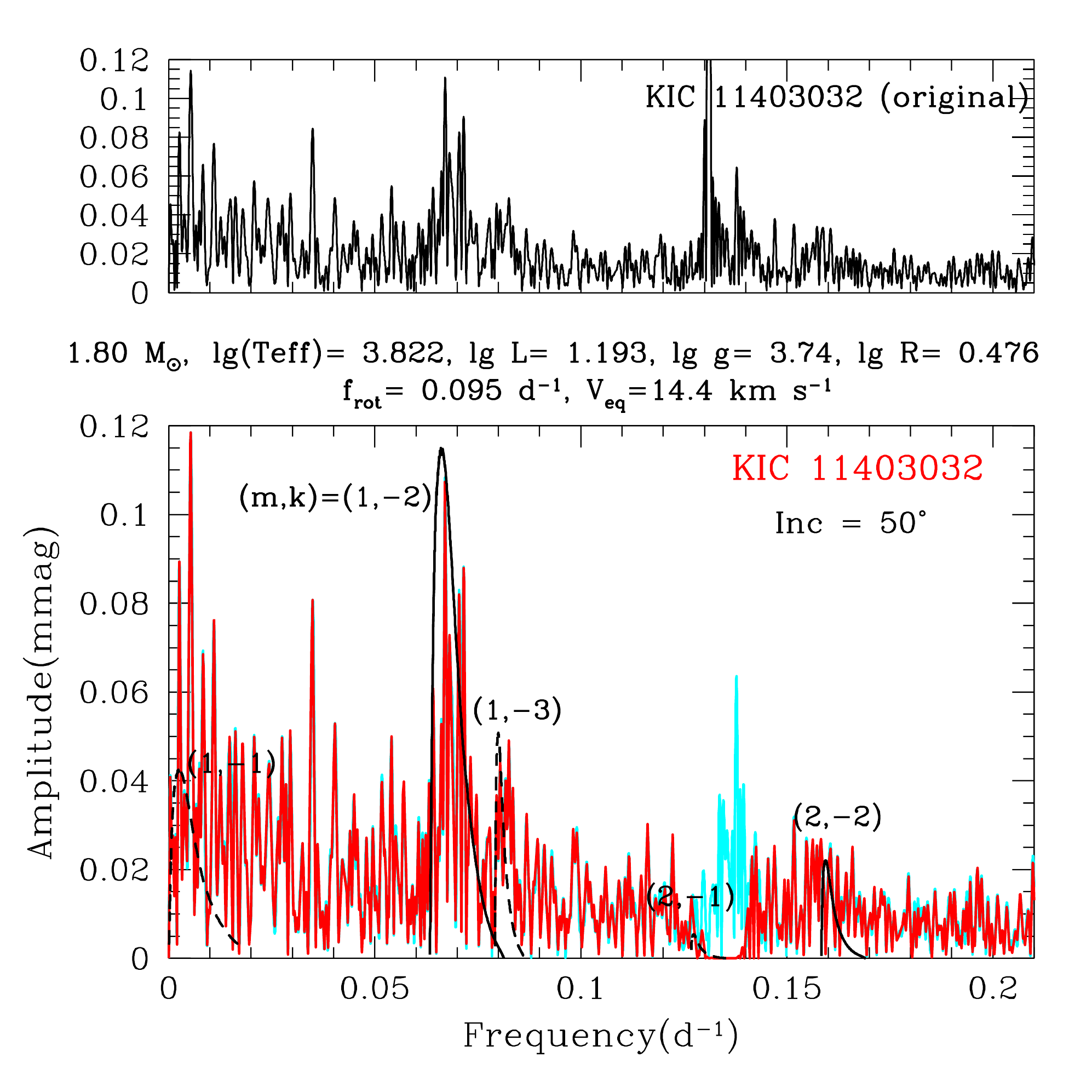}
\caption{FTs for KIC\,11403032 obtained from a KEBC \citep{KEBC2016} de-trended light curve. The upper panel shows an FT from the original data including the orbital frequency at $0.131$\,d$^{-1}$. The orbital frequency is removed in the lower panel. The cyan line denotes harmonics and combination frequencies of high-amplitude ($>0.05$~mmag) peaks around $0.07$\,d$^{-1}$. The group of peaks around $0.07$\,d$^{-1}$ is fitted with $(m,k)=(1,-2)$ r~modes in a 1.8-M$_\odot$ model at a rotation frequency of $0.095$\,d$^{-1}$. A narrow group at $\sim0.08$\,d$^{-1}$ agrees with with odd $(m,k)=(1,-3)$ r~modes. Furthermore, the relatively small amplitude group at $\sim$$0.16$\,d$^{-1}$ likely corresponds to $(m,k)=(2,-2)$ r~modes.}
\label{fig:k114}
\end{figure}

The orbital period of KIC\,11403032 is $7.6316$\,d (KEBC), while the eccentricity $e=0.288\pm0.013$ was obtained by \citet{Shporer2016}.   Equation~(\ref{eq:ps}) with these parameters gives $P_{\rm ps-rot}=5.05\pm0.22$\,d. The rotation period obtained by our r~mode fitting (Fig.\,\ref{fig:k114}) is $10.5$\,d ($f_{\rm rot}=0.095$\,d$^{-1}$), which is longer than $P_{\rm ps-rot}$ by about a factor of two. The rotation of KIC\,11403032 is the slowest among our sample. \citet{Zimmerman2017} obtained a still longer rotation period of $14.89\pm0.22$\,d for KIC\,11403032 assuming the frequency group at $\sim$$0.07$\,d$^{-1}$ in the FT to be caused by surface spots.

\subsection{KIC\,11568657}\label{sec:k115}

\begin{figure}
\includegraphics[width=0.49\textwidth]{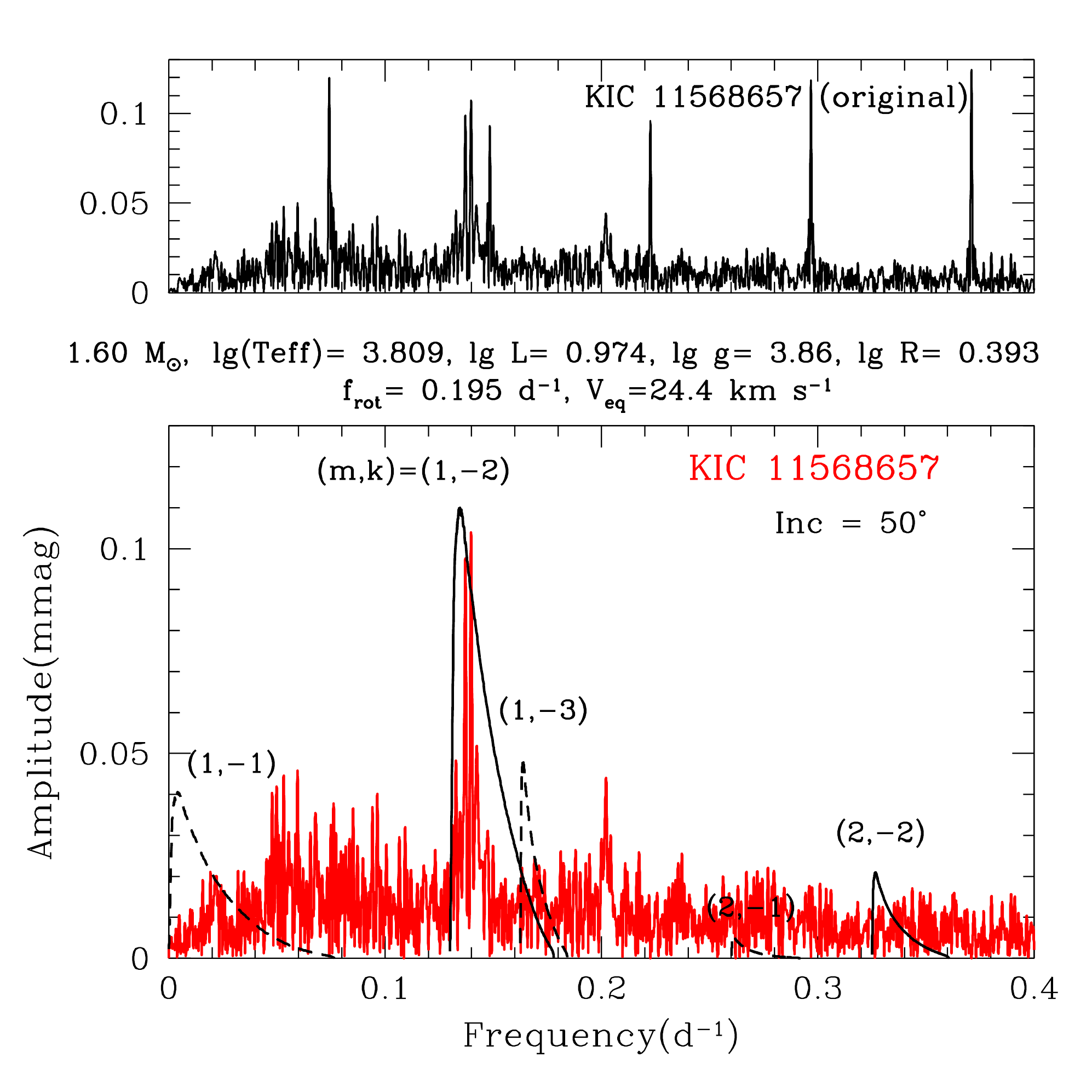}
\caption{FTs for the heartbeat star KIC\,11568657. The frequency group at $\sim$$0.14$\,d$^{-1}$ is fitted with $(m,k)=(1,-2)$ r~modes in a 1.6-M$_\odot$ model with a rotation frequency of $0.195$\,d$^{-1}$. The amplitude peak at $0.2~{\rm d}^{-1}$ is possibly caused by oscillatory convective modes in the convective core \citep{Lee2020,Lee2021}, or a spot on the surface.}
\label{fig:k115}
\end{figure}

The heartbeat star KIC\,11568657 has an orbital period of $13.476$\,d (KEBC). 
\citet{Thompson2012} obtained the orbital eccentricity $e= 0.565\pm0.002$ and inclination $i=48.2\pm 0.1^{\circ}$.  From the orbital period and the eccentricity, equation~(\ref{eq:ps}) yields  $P_{\rm ps-rot}=3.798\pm0.029$\,d, which is somewhat shorter than the rotation period $5.13$\,d obtained by the fitting the frequency group at $\sim0.14$\,d$^{-1}$ with $(m,k)=(1,-2)$ r~modes (Fig.\,\ref{fig:k115}). 

\subsection{KIC\,11649962}\label{sec:k116}

\begin{figure}
\includegraphics[width=0.49\textwidth]{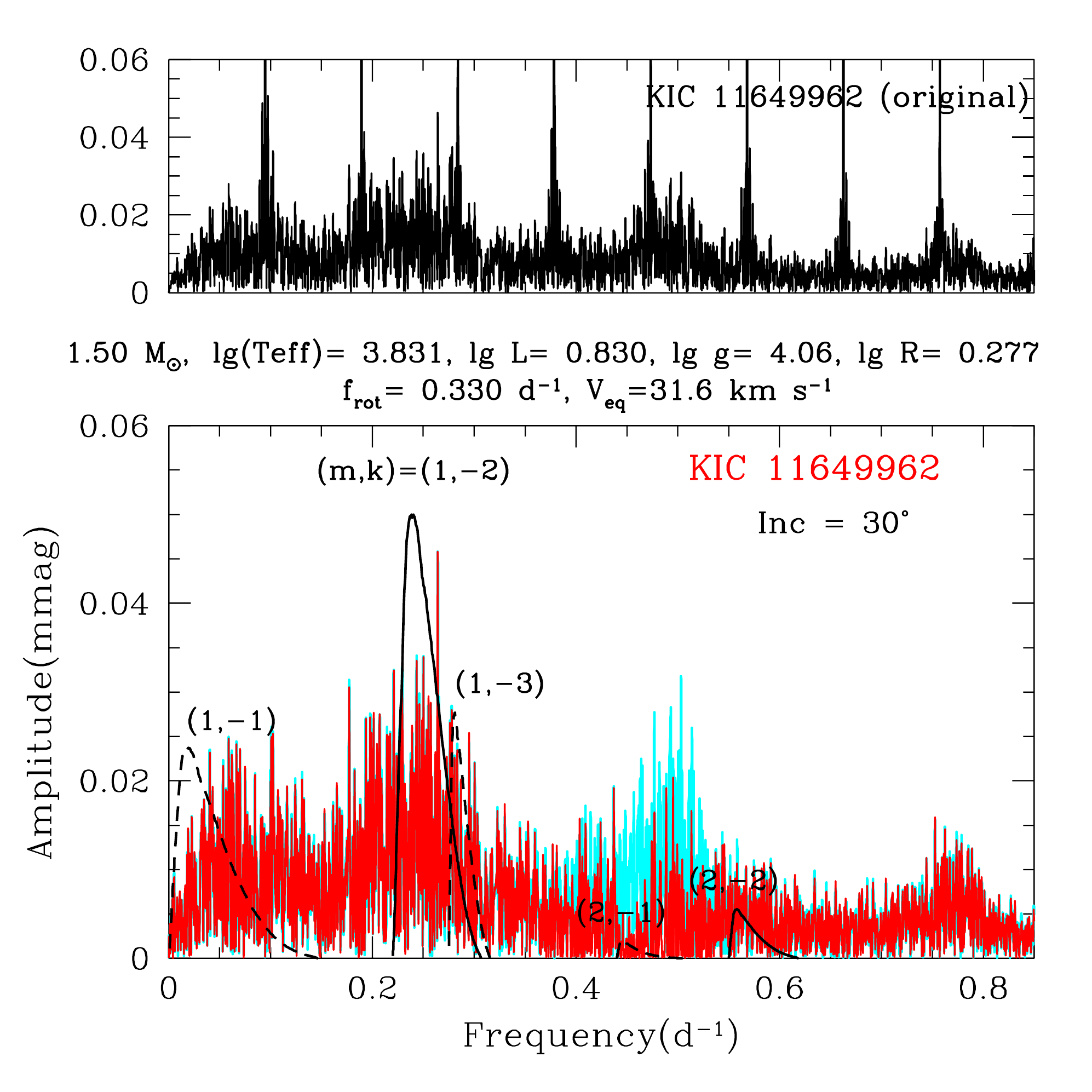}
\caption{FTs for  KIC\,11649962. The cyan line indicates harmonics and combination frequencies of fifteen highest amplitude ($>0.027$~mmag) frequencies around  $0.25$\,d$^{-1}$. The frequency group at $\sim$$0.25$\,d$^{-1}$ is fitted with the visibility curve of $(m,k)=(1,-2)$ r~modes in the 1.5-M$_\odot$ model at a rotation frequency of $0.33$\,d$^{-1}$.  }
\label{fig:k116}
\end{figure}

The heartbeat star KIC\,11649962 has an orbital period of $10.563$\,d (KEBC). \citet{Shporer2016} obtained the eccentricity $e= 0.5206\pm0.0035$. Using the orbital period and the eccentricity in equation~(\ref{eq:ps}) we obtain $P_{\rm ps-rot}= 3.505\pm0.043$\,d, which is comparable to the rotation period $3.03$\,d obtained by fitting the frequency group at $\sim$$0.25$\,d$^{-1}$ with the $(m,k)=(1,-2)$ r~modes (Fig.\,\ref{fig:k116}).

\subsection{KIC\,11923629}\label{sec:k119}

\begin{figure}
\includegraphics[width=0.49\textwidth]{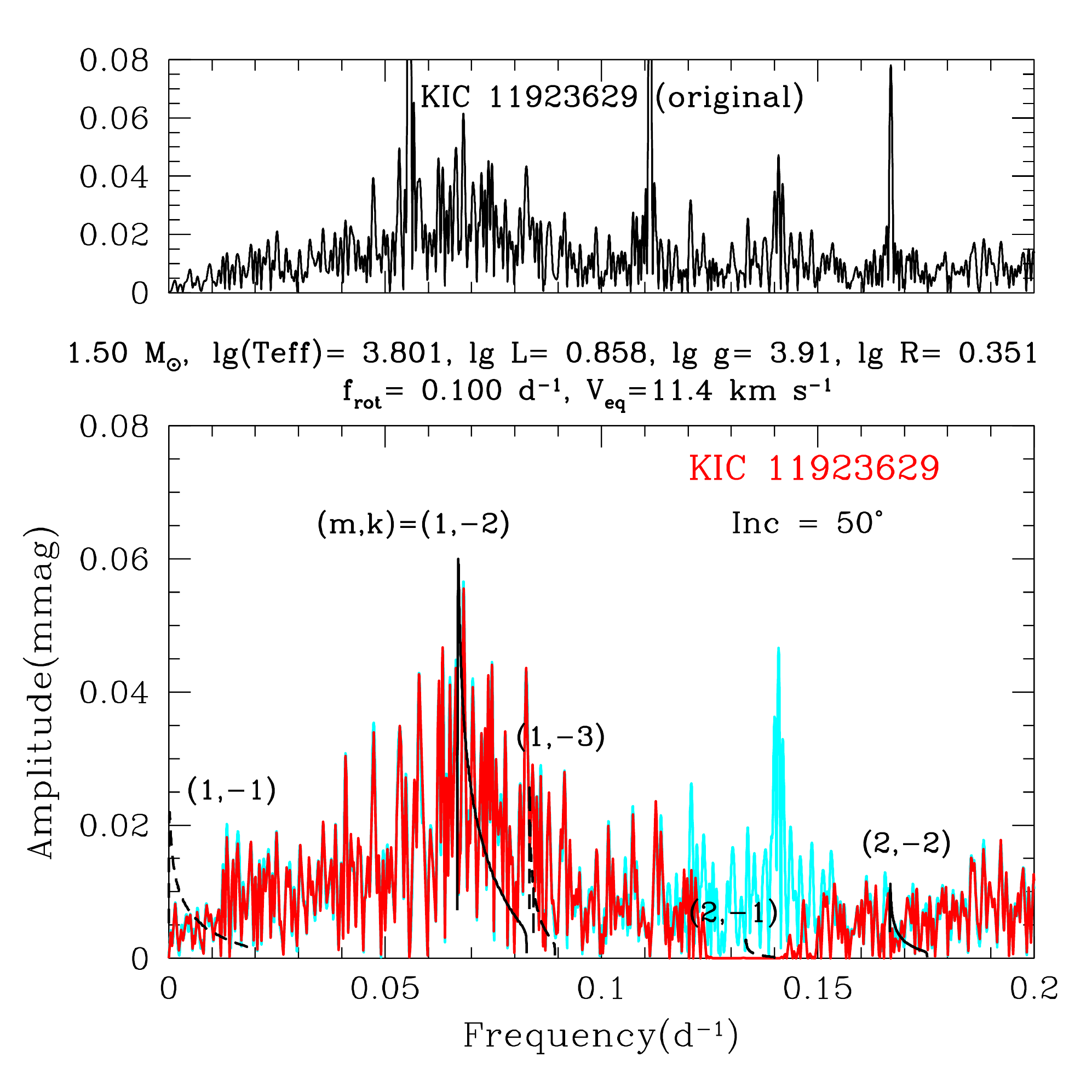}
\caption{FTs for  KIC\,11923629.  The cyan line denotes frequencies corresponding to harmonics and combination frequencies among highest amplitude ($>0.04$~mmag) frequencies around $0.07$\,d$^{-1}$. The frequency group around $0.07$\,d$^{-1}$ is fitted with $(m,k)=(1,-2)$ r~modes in a 1.50-M$_\odot$ model at a rotation frequency of $0.10 \pm 0.01$\,d$^{-1}$.}
\label{fig:k119}
\end{figure}

The heartbeat star KIC\,11923629 has an orbital period of $17.973$\,d (KEBC). \citet{Shporer2016} obtained the eccentricity $e=0.363\pm0.006$. From the orbital period and eccentricity we obtain $P_{\rm ps-rot}=9.77\pm0.16$\,d, which is comparable to the rotation period $10.0$\,d obtained by fitting with r-mode frequency range (Fig.\,\ref{fig:k119}), while \citep{Zimmerman2017} obtained a slightly longer rotation period of $15.79\pm0.15$\,d by assuming the frequency group at $\sim$$0.07$\,d$^{-1}$ to be generated by surface spots.

\subsection{KIC\,12255108}\label{sec:k122}

\begin{figure}
\includegraphics[width=0.49\textwidth]{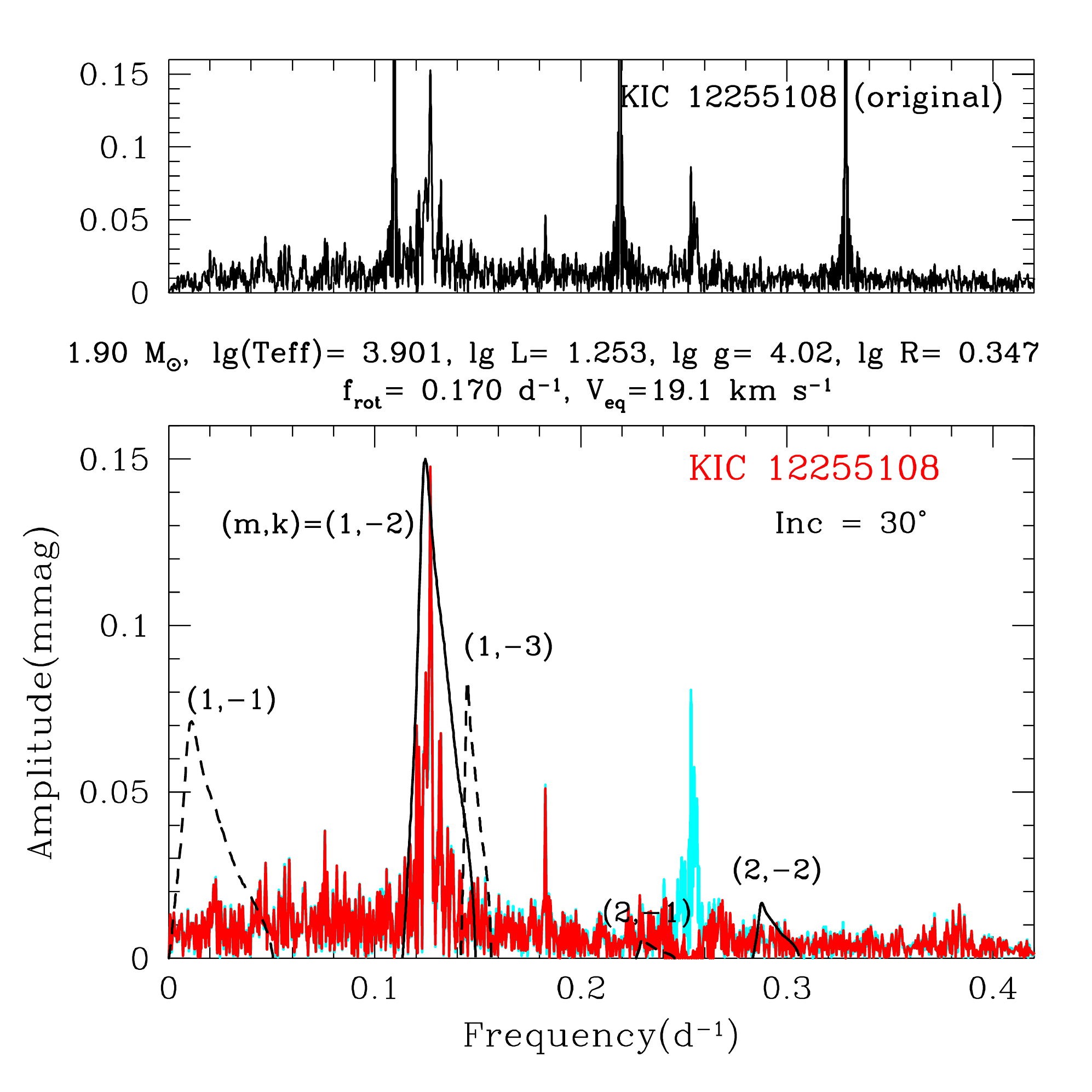}
\caption{FTs for KIC\,12255108. The cyan line denotes harmonics and combination frequencies of high amplitude peaks ($> 0.06$~mmag) around $0.12$\,d$^{-1}$.}
\label{fig:k122}
\end{figure}

The heartbeat binary KIC\,12255108 has an orbital period $P_{\rm orb}=9.1315$\,d (KEBC), and the orbital eccentricity $e=0.296$ obtained by \citet{Shporer2016}. Then, equation~(\ref{eq:ps}) gives the pseudo-synchronous period $P_{\rm ps-rot}= 5.92$\,d. Fig.\,\ref{fig:k122} (lower panel) shows an FT for KIC\,12255108 fitted with the visibility of r~modes at a rotation frequency of $0.17$\,d$^{-1}$ ($P_{\rm rot}=5.88$\,d) for a 1.90-M$_\odot$ main-sequence model. The obtained rotation period is similar to the pseudo-synchronous rotation period.

\bsp	
\label{lastpage}
\end{document}